\documentclass{aa} 
\usepackage{0343}
\usepackage{graphicx}
\begin{document}
   \title{The polarization of mm methanol masers}

   \author{H. Wiesemeyer \inst{1}\thanks{\email{wiesemey@iram.fr}}
          \and C. Thum\inst{1}
          \and
          C.M. Walmsley\inst{2}
          }

   \offprints{H. Wiesemeyer}

   \institute{Institut de Radio Astronomie Millim\'etrique, \\
              300, Rue de la Piscine, F-38406 Saint Martin d'H\`eres, France 
         \and
              Osservatorio Astrofisico di Arcetri, \\
              Largo E. Fermi, 5. 50125 Firenze, Italy}

   \date{Received 27 February 2004; accepted 27 July 2004}

   \abstract{
We present a survey of the polarization properties of mm-wavelength 
methanol masers, comprising both classes, and transitions from 84.5 to 
157.0\,GHz. Linear polarization is found in more than half of the observed 
objects, and circular polarization is tentatively detected in two sources.  
Class~I and Class~II \ch3oh  masers show similar polarization properties. 
The largest linear polarization is found in the 133\,GHz Class~I maser towards 
\object{L\,379} (39.5\,\%), and in the 157\,GHz Class~II maser towards 
\object{G\,9.62+0.19} (36.7\,\%). The spectral profiles of the polarization
angle of Class~I masers are mostly flat, except for two sources showing a linear
slope. Since the mm-line methanol masers are expected to be weakly (or not) 
saturated, we suggest that the stronger fractional polarizations found 
by us are enhanced by anisotropic pumping and radiative losses. In 
\object{NGC\,7538}, we find, for both maser classes, a good agreement between 
our polarization angles, and those measured for the submillimeter dust 
continuum. This can be taken as evidence for magnetic alignment of dust grains. 
It is also possible that an unsaturated maser with equally populated magnetic 
substates simply amplifies polarized continuum seed radiation. For Class~II
masers, the polarization properties of the various velocity components towards 
a given source with detectable polarization are quite homogeneous. A possible 
explanation is discussed. Since methanol is non-paramagnetic, the circular 
polarization of the unsaturated maser emission can only be due to variations 
of the angle between the magnetic field and the line of sight along the maser 
propagation path.

   }

   \maketitle
%

\section{Introduction}
The polarization of astronomical masers is a complex field of current research. 
Whereas the polarization properties of non--amplified radiation simply derive 
from asymmetries of the emitting volume, masers introduce additional aspects 
due to their excitation and propagation. In terms of excitation, methanol 
masers can be classified into two well distinguished classes\,: Class~I sources 
(prototype Orion BN/KL) are loosely associated with, but clearly offset from 
ultracompact H{\sc ii}--regions. They may form in outflows and cloud--cloud 
encounters. These masers are thought to be pumped by collisions into high 
rotational levels from where they decay spontaneously into the upper level of 
the masing transition (Cragg et al., \cite{1992MNRAS.259..203C}). Class~II 
sources (prototype W3\,OH) are often coincident with known OH masers and 
H{\sc ii}--regions, and may thus occur in the compressed envelopes exterior to 
the ionized gas. Their masing transitions suggest pumping by FIR photons to 
torsionally excited levels (Sobolev \& Deguchi, \cite{1994A&A...291..569S}). 
\\[1.5ex]
Only two studies of methanol maser polarization properties are published
(Caswell et al., \cite{1993MNRAS.260..425C}, Koo et al., 
\cite{1988ApJ...326..931K}). Both surveys were made at 12 GHz and cover only a 
small sample of Class~II masers. Linear polarization of up to 10\,\% towards 4 
(out of 11) and 2 (out of 5) sources respectively was detected in these two 
surveys. 
\\[1.5ex]
Here we present a survey of polarized millimeter-wave methanol masers, 
including both maser classes. The polarization of Class~I masers
was observed for the first time. The wealth of strong methanol masers detected 
at several mm transitions combined with the efficiency of a new polarimeter on 
the 30m telescope were instrumental in improving upon the two cm surveys 
with respect to the number of detections of polarization, the better 
polarization sensitivity, and the advantage of doing all Stokes parameters 
simultaneously. 

\section{Instrumentation and Observations}
The observations were done at the IRAM 30m telescope on Pico Veleta in February
and May 2002. We used the observatory's 3\,mm and 2\,mm receivers (brought into
phase with a common local oscillator reference) together with a versatile 
polarimeter working in the intermediate frequency band near 150\,MHz. The 
inherent frequency agility was essential in measuring the polarization of 
several transitions quasi simultaneously. The polarimeter (described in  
more detail in Thum et al., \cite{2003SPIE.4843..272T}) derives all four Stokes
parameters from the output signals of the two orthogonally polarized receivers,
after insertion of phase lags of $0^\circ$, $90^\circ$, $180^\circ$ and 
$270^\circ$. For each of these phase lags, the coherent superposition of the 
two signals yields an output signal whose power contains two terms, (i) the 
power from the two receivers, and (ii) a correlation term (like in an adding 
interferometer). The correlation terms are extracted by considering power 
differences between the four output channels, and yield Stokes I, U and V. 
Stokes~Q is derived from the difference of the total powers of the two 
receivers. Small phase offsets between both receivers (of order 
$\la 5^\circ/$hour) were regularly (typically once per hour) measured with a 
signal generator, and removed with a phase shifter acting on the local 
oscillator signal. The remaining instrumental polarization was regularly 
calibrated by means of unpolarized sources (mainly planets), for which all the 
output channels must yield the same signal strength. Any residual instrumental
polarization on the optical axis was thus removed. The remaining polarization 
sidelobes mainly arise from the optics in the receiver cabin, and are well 
known (Thum et al., \cite{2003SPIE.4843..272T}). For the observations
presented here, the beam polarization only limits the accuracy of the results 
for mispointed observations (see below), or masers spots scattered across a 
significant fraction of the polarization sidelobes. 
\\[1.5ex]
Since the polarimeter's six output signals were analysed 
by the facility autocorrelator, we were restricted to use a 39\,kHz 
channel spacing, together with a 17.5\,MHz bandwidth. We used a parabolic
apodization (Welch time lag window), yielding a half-power width of the
spectral response function of $1.59\times$ the channel separation, which is a 
good compromise between spectral resolution and sidelobe suppression. 
The resulting velocity resolutions are given in Table~\ref{maser_lines}, 
together with the characteristics of the observed maser lines. The observed 
linewidth was approximately deconvolved from the spectral response to the 
parabolic apodization, using a Gaussian fit to the main lobe of the latter. 
\\[1.5ex]
Our source selection is based on the strongest mm-wavelength masers found in the
available surveys. Class~I masers at mm wavelengths have been observed at 
84.5 \& 95\,GHz (Val'tts et al., \cite{1995ARep...39...18V}) and 133\,GHz 
(Slysh et al., \cite{1997ApJ...478L..37S}), mm Class~II masers 
at 107\,GHz (Val'tts et al., \cite{1995A&A...294..825V})
and 157\,GHz (Slysh et al., \cite{1995ApJ...442..668S}). 
Many of the positions of Class~II masers are from the 6.6\,GHz survey 
of Menten (\cite{1991ApJ...380L..75M}). Some mm methanol masers 
have been observed at the same velocities, but lower frequencies, with 
better positional accuracy (Kogan \& Slysh, 
\cite{1998ApJ...497..800K}, for 44\,GHz Class~I masers with the VLA, 
Minier et al., \cite{2000A&A...362.1093M}, for 6.7 \& 12.2 GHz Class~II masers
with the EVN and VLBA). 
\\[1.5ex]
In order to avoid false polarization detections, we retain only those masers 
that are not affected by the instrumental polarization sidelobes, applying the 
following criteria:
\\[1.5ex]
(1) Any mispointing results, even for an unpolarized signal, in an 
instrumental polarization. Since the latter is constant across the intermediate
frequency band, a scaled copy of the Stokes\,I signal would appear in 
Stokes\,Q, U and V. In that case, the scaling factors must be compatible with 
the instrumental signature, known from our polarization sidelobe measurements.
If this is the case, those data are discarded. For two or more masers offset 
within the beam, the result is not unique anymore, due to the convolution 
with the polarization sidelobes.
\\[1.5ex]
(2) Wherever interferometric observations of mm- and other methanol masers 
indicate that the maser spots are scattered across the polarization sidelobes 
of the optical system of the 30\,m telescope, we discard the data. This is the 
case e.g. for the two 84\,GHz maser components in DR21(OH) (Batrla \& Menten, 
\cite{1988ApJ...329L.117B}).
\\[1.5ex]
The linear polarization given here is 
\begin{equation}
p_{\rm L} = \frac{
\sqrt{Q_{\rm line}^2+U_{\rm line}^2}}{I_{\rm line}}
\end{equation}
i.e. the linear polarization is quoted against the spectral line Stokes\,I 
flux density (after subtraction of a contribution from the continuum emission).
The relative error of the fractional linear polarization is (to first order in 
$p_{\rm L}$) given by the signal-to-noise ratio of the Stokes I measurement. 
Continuum polarization measurements, as compared to spectral line polarimetry, 
are thus more precise, owing to the larger bandwidth. A few sources display 
continuum emission sufficiently strong to contribute to the signal in our quite
narrow spectral band. The importance of dust polarization at mm wavelengths is 
still a matter of debate. Linear polarization (up to 8\,\%) of the dust 
continuum at $\lambda\,1.3$\,mm was observed towards the Orion cloud core, 
offset from the BN/KL position (Leach et al., \cite{1991ApJ...370..257L}). 
Towards BN/KL, the fractional polarization drops to 2.7\,\% ("polarization
hole"). In our case, a mispointing of $10''$ can also lead to a linear 
polarization of up to 8\,\% (depending on the direction of the mispointing). The
fractional continuum polarization may also be affected by the measurement 
inaccuracy of Stokes~I due to fluctuations of the atmospheric emission (the IF 
polarimeter is not designed to measure weak continuum polarization levels). 
Stokes Q, U and V are not concerned, since they result from differential and 
correlation measurements, respectively, so our continuum polarization angle 
measurement remains reliable - on condition that the continuum emission is 
unresolved. 

The statistical bias, introduced 
in the determination of $\sqrt{Q^2+U^2}$, was subtracted from the percentage polarization (see Wardle \& Kronberg, 
\cite{1974ApJ...194..249W}). The standard deviations of the linear polarization
and its position angle were derived from the polarimeter's output signals
by error propagation.

\begin{table*}
\caption[]{The observed methanol masers. All masers are in the torsional ground
state. Rest frequencies are from Anderson et al. (\cite{1990ApJS...72..797A}).}
      \label{maser_lines}
$$
         \begin{array}{rrrrrlrcccc}

            \hline
            \hline
            \noalign{\smallskip}
            \mathrm{J_u} & \mathrm{K_u} & & \mathrm{J_l} & 
            \mathrm{K_l} & \mathrm{Symm.} & \mathrm{Frequency} & 
            \mathrm{E_u} & \mathrm{E_l} & \mathrm{Vel. resolution} &\mathrm{Maser} \\
                         &              & &              & 
                         & \mathrm{Substate} & \,\mathrm{[MHz]}\hfill     & 
            \mathrm{[cm^{-1}]} & \mathrm{[cm^{-1}]} & \mathrm{[km\,s^{-1}]} & \mathrm{class} \\
            \noalign{\smallskip}
            \hline
            \noalign{\smallskip}
             5 & -1 & \rightarrow & 4 & 0  & E   & 84521.210 & 27.417  & 24.597 & 0.221 & (I) \\
             8 &  0 & \rightarrow & 7 & 1  & A^+ & 95169.440 & 14.905  & 11.705 & 0.196 & (I) \\
             3 &  1 & \rightarrow & 4 & 0  & A^+ & 107013.850 & 19.704 & 16.134 & 0.173 & (II) \\
             6 & -1 & \rightarrow & 5 & 0  & E   & 132890.790 & 37.092 & 32.660 & 0.140 & (I) \\
             6 &  0 & \rightarrow & 6 & -1 & E   & 157048.620 & 42.331 & 37.092 & 0.119 &(II) \\
            \noalign{\smallskip}
            \hline
         \end{array}
$$
\end{table*}

\begin{table*}
      \caption[]{Detections ($> 3\sigma_{\rm rms}$) of linear polarization 
towards Class~I \ch3oh masers. Only velocity channels with the most significant
polarization are given here (with velocities in column 10). For the other 
velocities, see Figs.~\ref{spectra_84ghz} to \ref{spectra_157ghz}). The maser 
polarization is given in columns 8 (fractional linear polarization $p_{\rm L}$)
and 9 (polarization angle $\chi$, East from North). The maser fluxes and 
half-maximum full widths are derived from least-square fits of Gaussian 
components; the resulting linewidths are corrected for the instrumental 
broadening (for details see text).
\label{maser_survey_class1}} 

      \begin{tabular}{llllrlcrrr}
            \hline
            \hline
            \noalign{\smallskip}
Source & Ref.  & \ra      
               & \dec~~~~~~~~ 
               & Frequency & \Fnu & $\Delta v_{\rm FWHM}$ 
               & \pl       & $\chi$ & $v_{\rm lsr}$ \\
       &       & 
               & 
               & [GHz]     & [Jy] & [\kms]  
               & [\%]      & [$^\circ$] & [\kms] \\
            \noalign{\smallskip}
            \hline
            \noalign{\smallskip} \\

\object{OMC-2}          & (1) & $05^{\rm h}35^{\rm m}27\fs07$ 
               & $-05\degr 09\arcmin\,52\farcs5$
               & 132.891 & 37.7 & 0.36 
& $  8.94 \pm 0.91 $ & $  22.1  \pm 2.6 $ & 11.21 \\ 
\object{S\,231}         & (2) & 05\,\,\,39\,\,\,\,13.06 &  +35\, 45\, 51.3 
               &  95.169 & 12.7 &  0.51 
& $  3.79 \pm 0.58 $ & $ 118.1  \pm 4.5 $ & -16.82 \\ 
\object{NGC\,2264}     & (1) & 06\,\,\,41\,\,\,\,08.02 &  +09\, 29\, 40.0 
               & 132.891 & 14.0 & 0.62 
& $ 10.6  \pm 1.7  $ & $ 106.4  \pm 4.1 $ & 7.22 \\ 
\object{M8E}            & (2) & 18\,\,\,04\,\,\,\,53.76 & $-$24\, 26\, 35.3 
               & 132.891 & 24.6 & 0.63 
& $ 33.2  \pm 1.2  $ & $ 154.56 \pm 0.8 $ & 10.93 \\   
\object{W\,33-Met}      & (1) & 18\,\,\,14\,\,\,\,11.08 & $-$17\, 55\, 57.4 
               & 132.891 & 8.8 & 0.79  
& $ 22.8  \pm 3.0  $ & $ 154.5  \pm 3.4 $ & 32.95 \\
\object{L\,379}         & (1) & 18\,\,\,29\,\,\,\,24.69 & $-$15\, 15\, 19.0 
               & 132.891 & 3.8 & 0.36  
& $ 39.5  \pm 6.0  $ & $ 133.7  \pm 3.4 $ & 19.18 \\ 
\object{W\,51-Met\,2}     & (3) & 19\,\,\,23\,\,\,\,46.50 &  +14\, 29\, 41.0 
               &  84.521 & 6.0 & 0.38  
& $  9.2  \pm 2.0  $ & $  23.6  \pm 5.6 $ & 56.12 \\ 
               &      &      &
               & 132.891 & 3.0 & 0.49 
& $ 20.1  \pm 1.1  $ & $  10.2  \pm 3.4 $ & 53.28 \\ 
\object{DR\,21-W}       & (1) & 20\,\,\,38\,\,\,\,54.72 &  +42\, 19\, 22.4 
               &  84.521 & 85.6 & 0.35 
& $  2.77 \pm 0.15 $ & $  48.1  \pm 1.4 $ & -3.15 \\ 
               &      &      &
               & 132.891 & 72.5 & 0.42 
& $  4.26 \pm 0.50 $ & $   7.7  \pm 3.0 $ & -2.44 \\ 
\object{W\,75 S(3)}     & (3) & 20\,\,\,39\,\,\,\,03.50 &  +42\, 25\, 53.0 
               &  84.521 & 14.3 & 0.28 
& $  9.4  \pm 2.0  $ & $   7.8  \pm 5.7 $ & -3.62 \\ 
\object{NGC\,7538-Met1} & (3) & 23\,\,\,13\,\,\,\,46.40 &  +61\, 27\, 33.0 
               &  84.521 & 3.8 &  0.70 
& $  5.5  \pm 1.6  $ & $ 161.2  \pm 5.0 $ & -58.00 \\
               &      &      & 
               &  95.169 & 14.1 &  0.82 
& $ 14.5  \pm 1.3  $ & $   7.2  \pm 2.6 $ & -57.42 \\
               &      &      & 
               & 132.891 & 3.5 &  0.81 
& $ 24.6  \pm 1.8  $ & $  53.3  \pm 1.9 $ & -57.35 \\

            \noalign{\smallskip}
	    \hline
      \end{tabular}
~\\[1.0ex] 
References: (1) VLA positions of the 44\,GHz maser, Kogan \& Slysh 
(\cite{1998ApJ...497..800K}), (2) VLBI positions of the 6.7 \& 12.2\,GHz masers,
Minier et al. (\cite{2000A&A...362.1093M}), (3) positions of the 95\,GHz survey
of Val'tts (\cite{1995ARep...39...18V}).
\end{table*}
\begin{table*}
      \caption[]{As Table\,\ref{maser_survey_class1}, but for Class~II \ch3oh 
masers. The distinct main velocity components are listed separately.
\label{maser_survey_class2}}
      \begin{tabular}{llllrlcrrr}
            \hline
            \hline
            \noalign{\smallskip}
Source          & Ref. & \ra      
                & \dec~~~~~~~~ 
                & Frequency & \Fnu & $\Delta v$ 
                & \pl       & $\chi$  & $v_{\rm lsr}$ \\
                &      & 
                & 
                & [GHz]     & [Jy] & [\kms]
                & [\%]      & [$^\circ$]    & [\kms] \\
            \noalign{\smallskip}
            \hline
            \noalign{\smallskip} \\

\object{NGC\,7538-IRS1}  & (1) & $23^{\rm h}13^{\rm m}45\fs36$ 
                & $+61\degr 28\arcmin09\farcs7$ 
                & 107.014 & $< 16.3$ & $< 2.34$  
& $  3.63 \pm 0.97 $  & $ 104.5 \pm 6.8 $ & -56.30 \\
                &     &    & 
                & 107.014 & $< 11.8$ & $< 1.57$ 
& $  5.7  \pm 1.2  $  & $ 116.2 \pm 5.8 $ & -58.59 \\
                &     &    & 
                & 107.014 & $ 8.9 $ & 0.88 
& $  6.1 \pm 1.6 $  & $ 112.6 \pm 6.7 $ & -60.56 \\
\object{Cep\,A}          & (1) & 22\,\,\,56\,\,\,\,18.10 & $+$62\, 01\, 49.5 
                & 157.049 & 19.5 & 0.48  
& $ 12.3  \pm 1.7  $  & $  86.4 \pm 4.0 $ & -3.84 \\ 
                &     &     & 
                & 157.049 & 40.6 & 0.34 
& $ 14.10 \pm 0.94 $  & $  87.2 \pm 2.0 $ & -2.80 \\ 
                &     &     &
                & 157.049 & 17.0 &  0.57 
& $ 14.0  \pm 2.0  $  & $  79.0 \pm 4.2 $ & -1.75 \\ 
                &     &     &      
                & 157.049 & 10.2 &  0.58 
& $ 10.6  \pm 2.7  $  & $  81.2 \pm 6.9 $ & -1.30 \\ 
\object{G\,9.62+0.19}    & (1) & 18\,\,\,06\,\,\,\,14.80 & $-$20\, 31\, 32.0 
                & 157.049 & 6.4& 0.88 
& $ 31.5  \pm 3.7  $  & $ 142.9 \pm 3.1 $ & 1.55 \\ 
                &     &     & 
                & 157.049 & 3.5 &  0.66  
& $ 36.7  \pm 5.7  $  & $ 135.1 \pm 4.8 $ & -0.09 \\ 
            \noalign{\smallskip}
	    \hline
      \end{tabular}
~\\[1.0ex] 
Note: The masers of \object{NGC\,7538-IRS1} are blended with thermal emission, 
only upper limits can be given for flux and linewidth.\\
References: (1) VLBI positions of Minier et al. (\cite{2000A&A...362.1093M}) 
\end{table*}
\section{Results}
\begin{figure}[h!]
   \resizebox{\hsize}{!}{\includegraphics{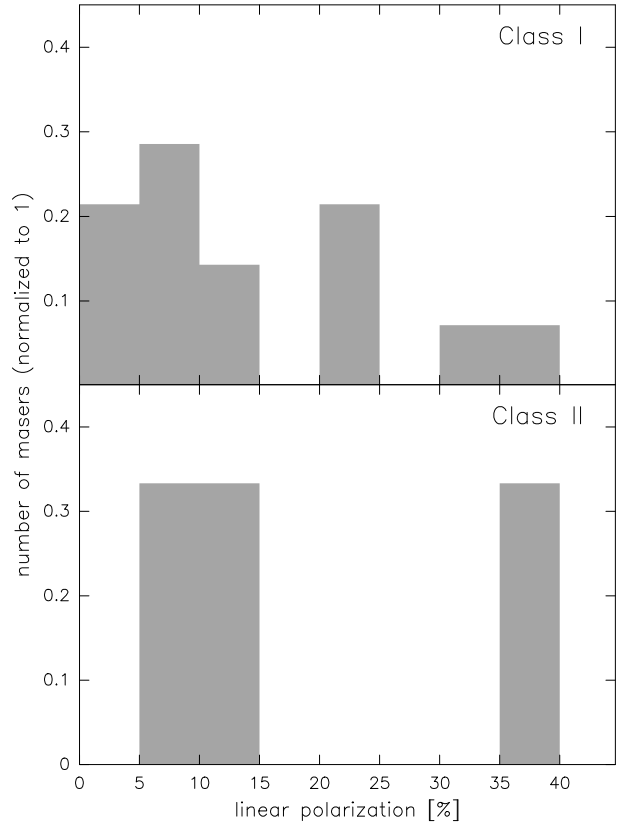}} 
   \caption{The distribution of detections of linear polarization towards
Class~I masers (top, for a total of ten sources) and Class~II masers (bottom, 
for three sources), binned into 5\,\% intervals. 
\label{histograms}}
   \end{figure}

In the following, the detections of methanol maser polarization are
discussed within the context of the environment of their star-forming 
regions. The two maser classes are well separated, i.e. each of the transitions 
summarized in Table~\ref{maser_lines} show maser action only in one of the two 
\ch3oh  maser classes. We find (all transitions confounded) polarization in 10 
out of 14 Class~I sources satisfying our selection criteria, and in 3 out 
of 7 Class~II sources. We cannot break down this statistics into specific 
transitions, since not all sources were observed at all frequencies of 
Table~\ref{maser_lines}. The distribution of polarization detections is shown 
in Figure\,\ref{histograms}. The detections of linear polarization range 
between 2.8\,\% and 39.5\,\%. The velocity profiles of polarization angles are 
either flat, or show a linear slope. Circular polarization is tentatively 
($\ga 3\,\sigma_{\rm rms}$) detected towards two Class~I masers. Our results 
are summarized in Table\,\ref{maser_survey_class1} for Class~I mm \ch3oh  
masers, respectively in Table\,\ref{maser_survey_class2} for their Class~II 
counterparts.
Given the non-uniformity of our data set, and the fact that the accuracy of the 
polarization measurements depends on the signal-to-noise ratio, the upper limits
for the non-detections depend much on the frequency and the weather conditions.
Typical upper limits for non-detections range from $p_{\rm L} = 0.5\,\%$ to 
about 10\,\%.

\subsection{Class~I Masers}

\begin{figure*}
   \centering
     \resizebox{8.5cm}{!}{\includegraphics*[bb=70 120 470 730]{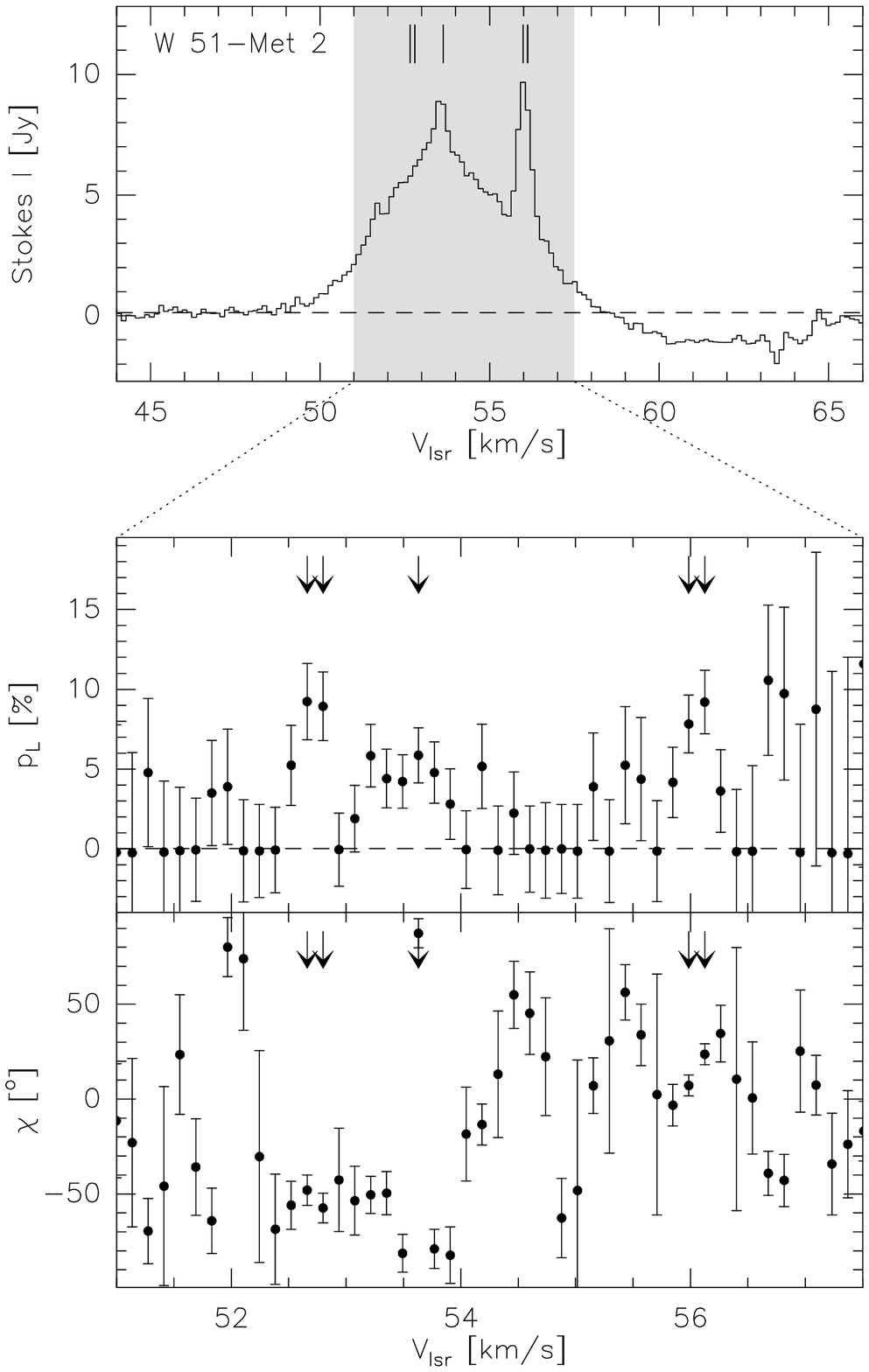}}
     \resizebox{8.5cm}{!}{\includegraphics*[bb=70 120 470 730]{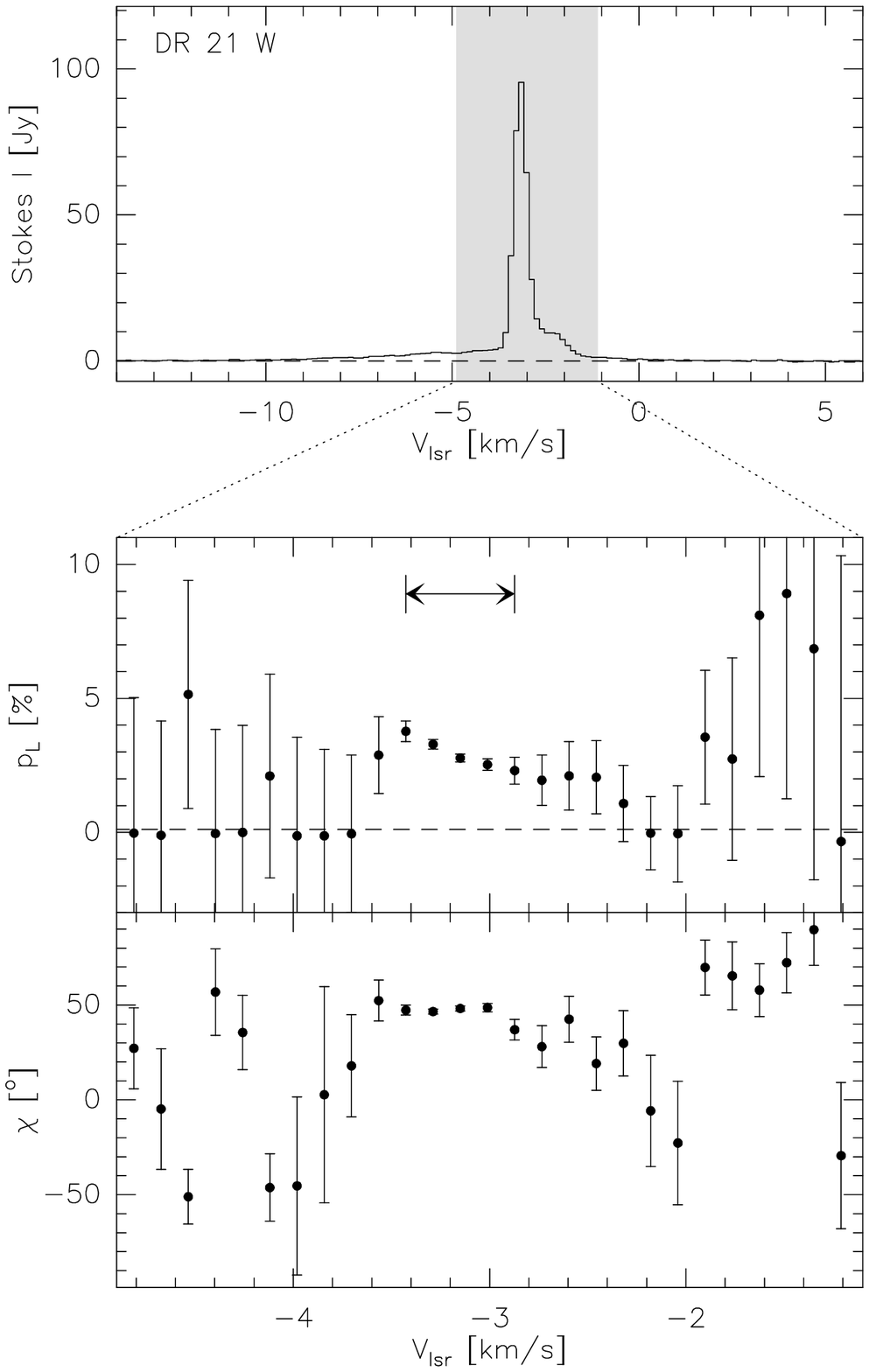}}
     \resizebox{8.5cm}{!}{\includegraphics*[bb=70 120 470 730]{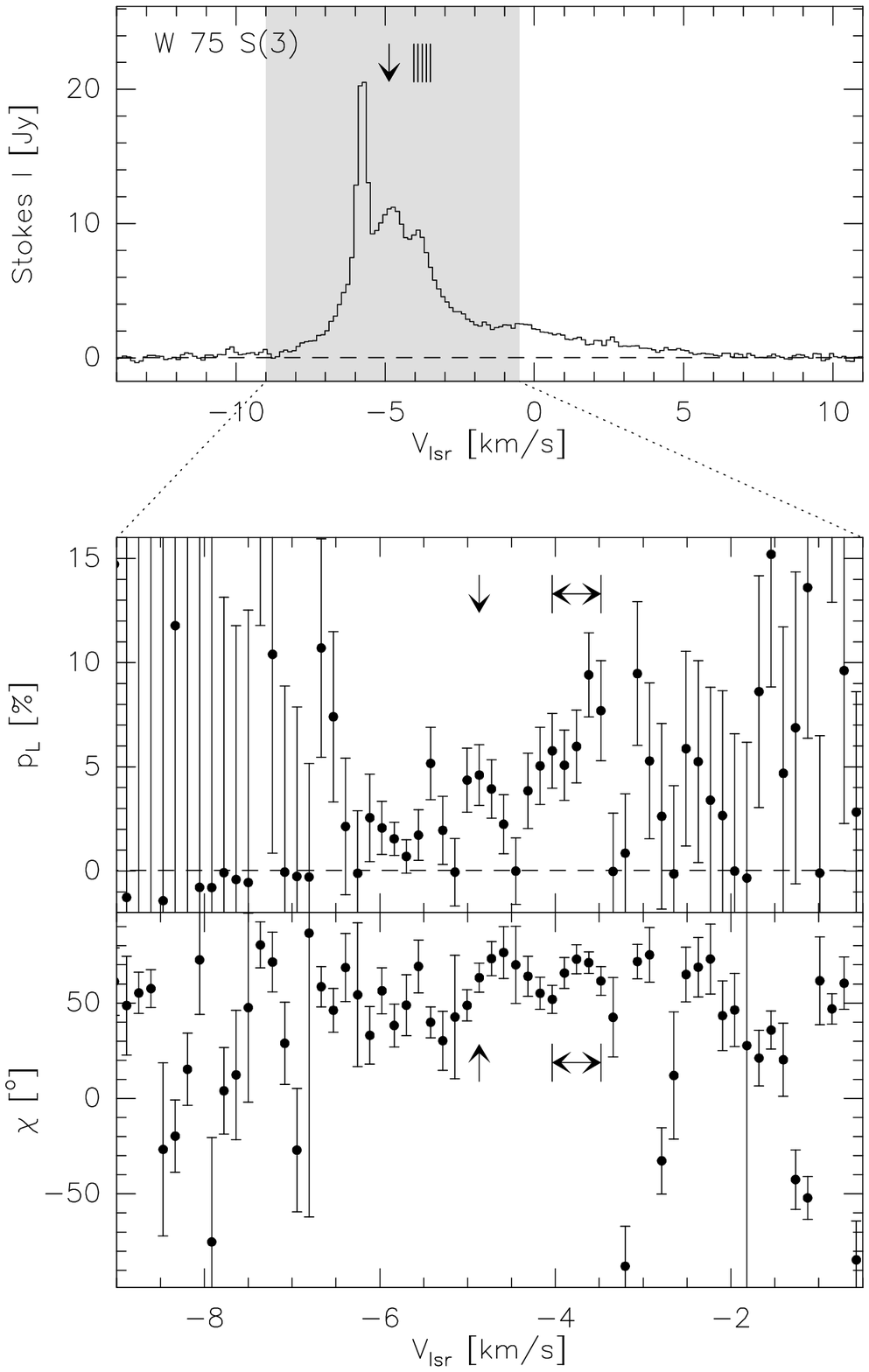}}
     \resizebox{8.5cm}{!}{\includegraphics*[bb=70 120 470 730]{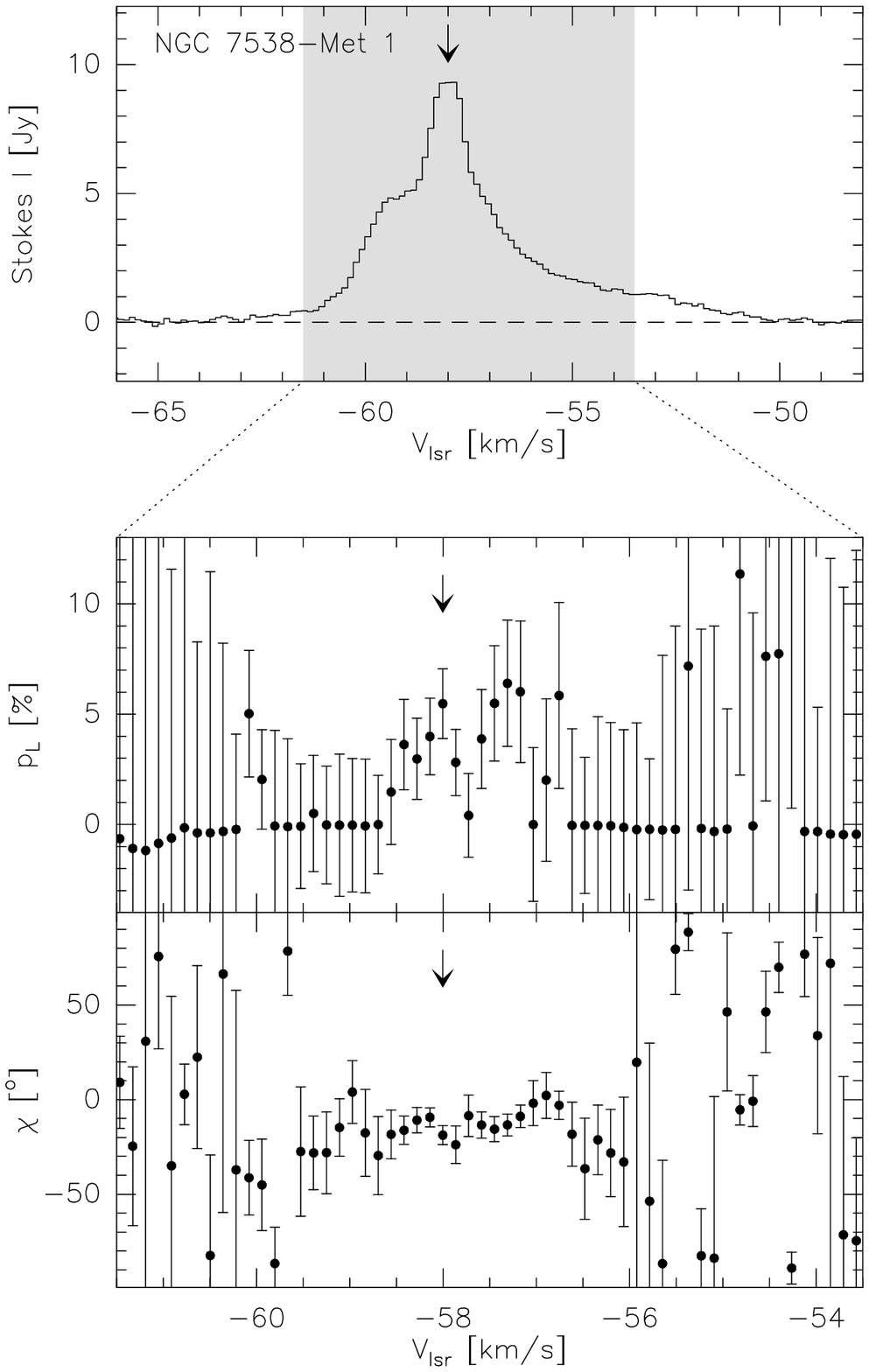}}
     \caption{Class~I methanol masers at 84\,GHz. For each source, 
     Stokes~I (top), the fractional linear polarization $p_{\rm L}$ (center), 
     and the polarization angle $\chi$ (E from N, bottom) are shown, with 
     spectral baselines subtracted. Vertical bars or arrows and horizontal 
     arrows mark the velocities or velocity ranges, respectively, with 
     $p_{\rm L} > 3\sigma_{\rm rms}$. The grey-shaded areas in the Stokes~I 
     spectra indicate the velocity range for $p_{\rm L}$ and $\chi$.
     \label{spectra_84ghz}}
   \end{figure*}

\begin{figure*}
   \centering
     \resizebox{8.5cm}{!}{\includegraphics*[bb=70 120 470 730]{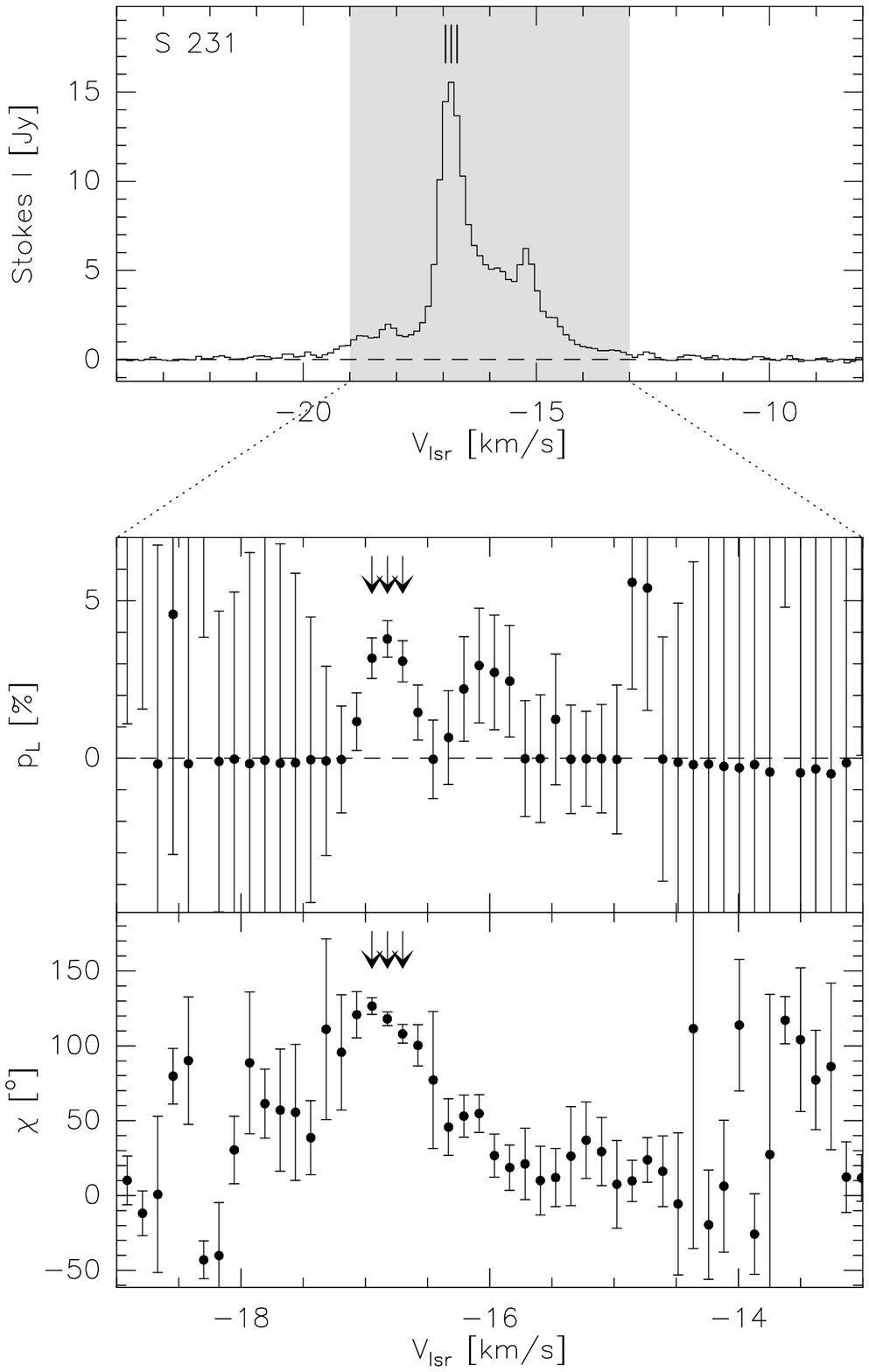}}
     \resizebox{8.5cm}{!}{\includegraphics*[bb=70 120 470 730]{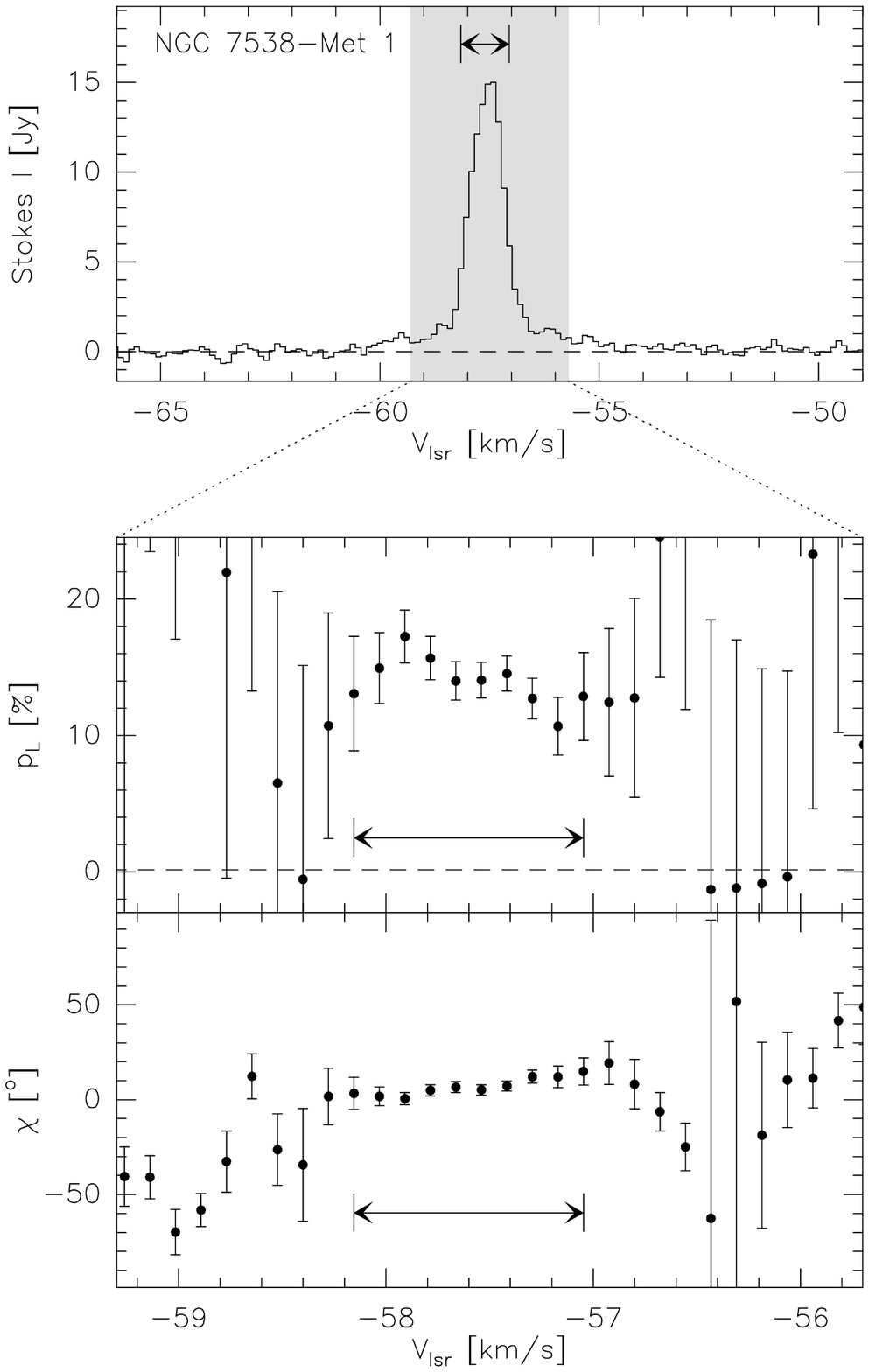}}
     \caption{Class~I methanol masers at 95\,GHz.\label{spectra_95ghz}}
   \end{figure*}

The polarization spectra for Class~I masers are shown in
Figures\,\ref{spectra_84ghz} to \ref{spectra_133ghz}, for 84\,GHz, 
95\,GHz, and 133\,GHz, respectively. The spectra of the fractional linear
polarization (in percent) and its position angle are for the velocity range 
indicated by the grey-shadowed area in the Stokes I spectra. All polarization 
angles are given in counterclockwise sense (E from N).
 
\subsubsection{\object{OMC-2}} 
We observed this maser, located in a cluster of infrared sources and
far-infrared condensations with multiple outflow activity, in the 133\,GHz 
$6_1 \rightarrow 5_0$ transition of E-type methanol (first detection by Slysh 
et al., \cite{1997ApJ...478L..37S}). Other Class~I transitions were previously
observed, but with unreliable astrometry. The maser peaks at 11.35\,\kms,  
the strongest polarization ($p_{\rm L} = 8.9$\,\%) is observed at 
11.21\,\kms. At first sight, the profile of the linear fractional polarization
looks like a scaled copy of the Stokes I line profile. However, the polarized
feature is significantly narrower than the maser component (0.37\,\kms vs. 
0.69\,\kms, respectively). An instrumental origin can also be ruled out, since
criterion (1) of section 2 does not apply. The polarization angle ($22^\circ$) 
remains constant across the line profile. As a matter of fact, the outflow 
driven by the source FIR\,3 and mapped by Williams et al. 
(\cite{2003ApJ...591.1025W}) has a position angle of 
$30^\circ$; the brightest maser spots of the $7_0\rightarrow 6_1\,A^+$ \ch3oh  
maser (at 44\,GHz, mapped with the VLA by Kogan \& Slysh, 
\cite{1998ApJ...497..800K}) are on a straight line of position angle 
$32^\circ$ (E from N). Both position angles are close to our polarization 
angle. This interesting finding, schematically shown in Figure~\ref{omc2}, will
be discussed below. The position and velocity of the maser spike corresponds to
one of the components of the 44\,GHz maser (component number 7 of Kogan \& 
Slysh). The thermal linewings make it impossible to know whether we also get 
weaker maser emission at the velocities of the other VLA maser spots, since the 
velocity dispersion of the latter is merely $0.23$\,\kms. If so, they could 
produce an important instrumental polarization, because they are spread over a 
N-S line extending over $15"$, i.e. as large as the telescope's main beam at 
that frequency. However, this would not be compatible with our flat 
polarization angle profile, since the sidelobe polarization pattern is 
significantly different in Stokes Q and U (see Thum et al.,
\cite{2003SPIE.4843..272T}).

   \begin{figure*}
   \centering
     \resizebox{8.5cm}{!}{\includegraphics*[bb=70 120 470 730]{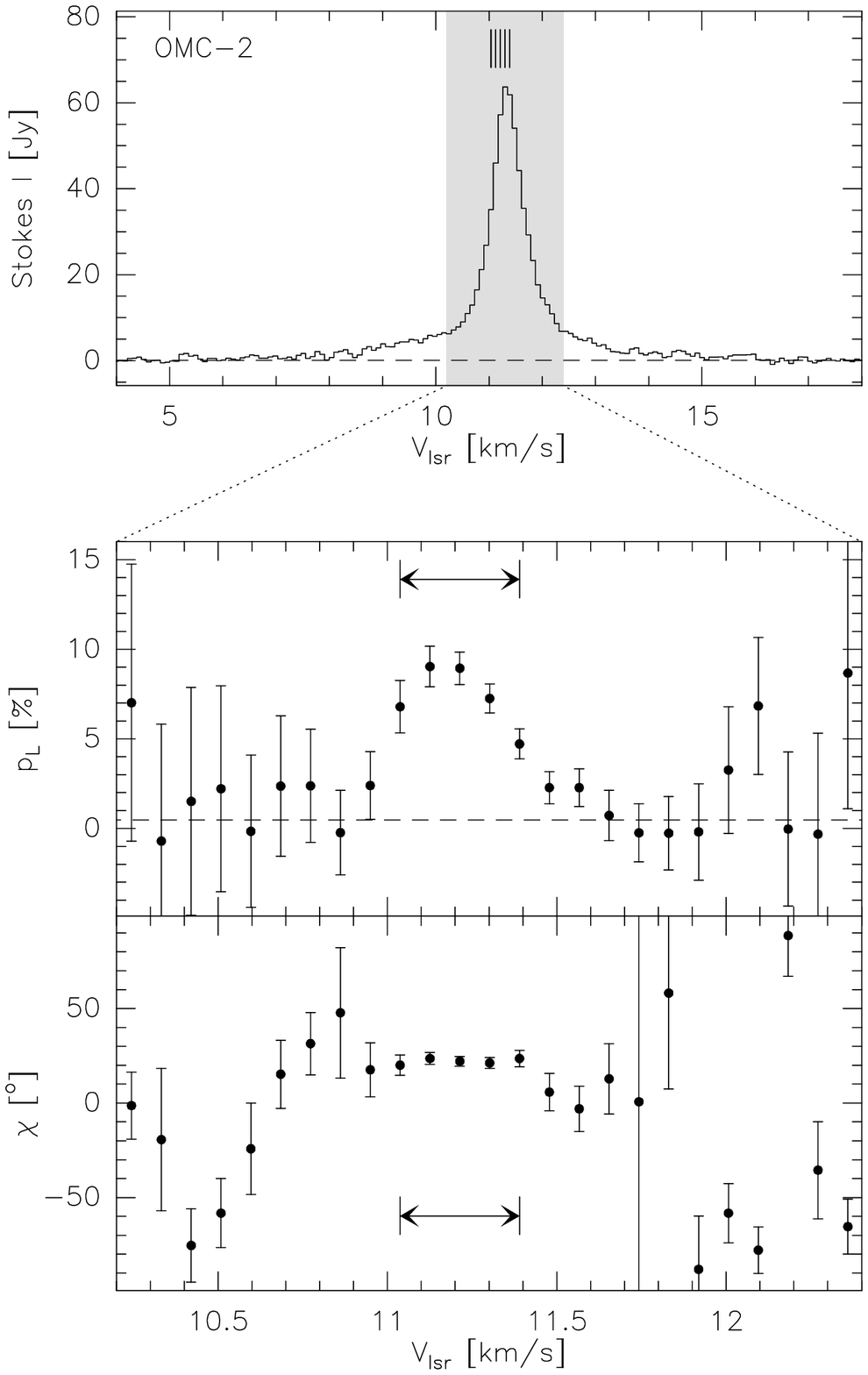}}
     \resizebox{8.5cm}{!}{\includegraphics*[bb=70 120 470 730]{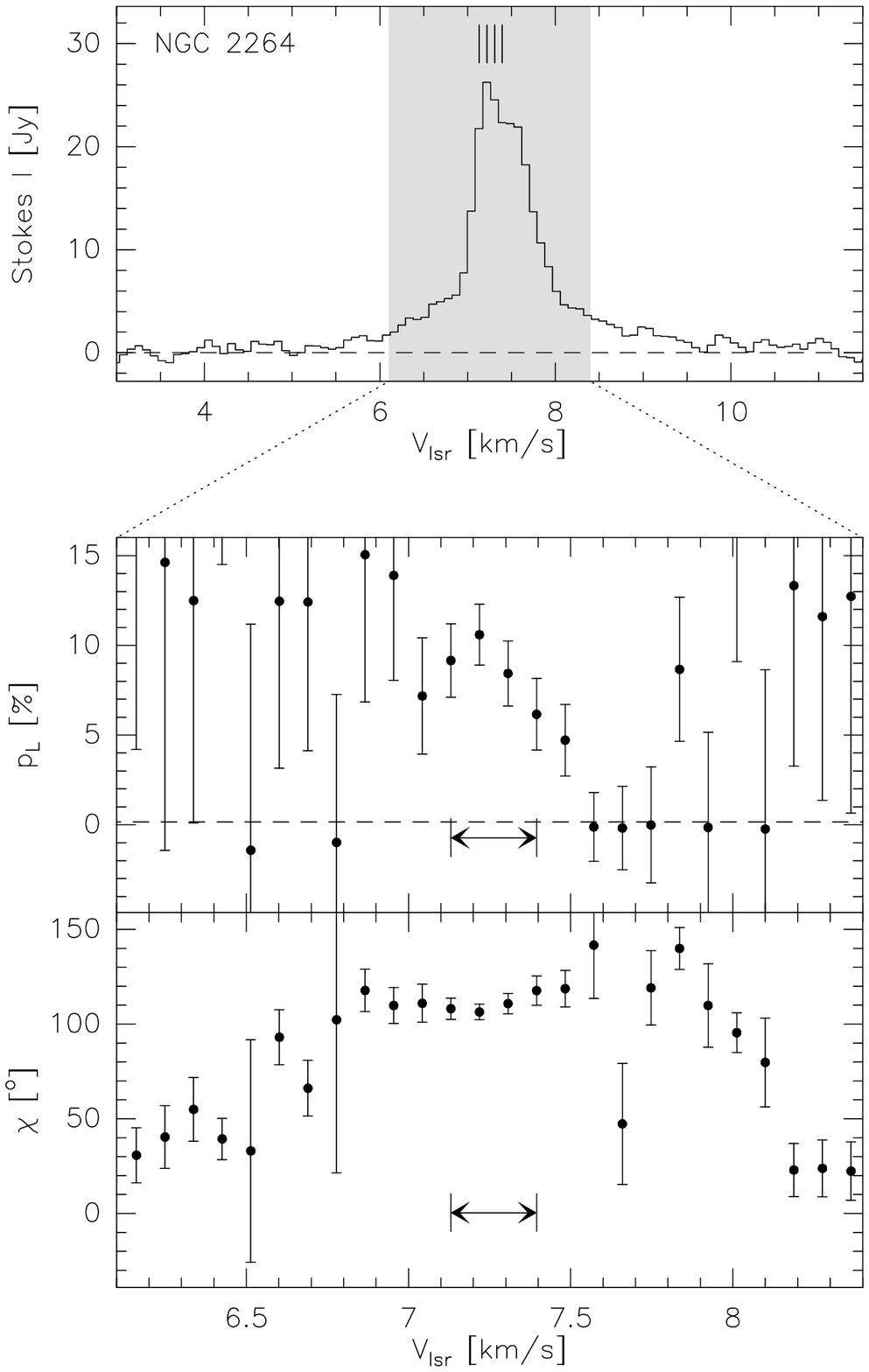}}
     \resizebox{8.5cm}{!}{\includegraphics*[bb=70 120 470 730]{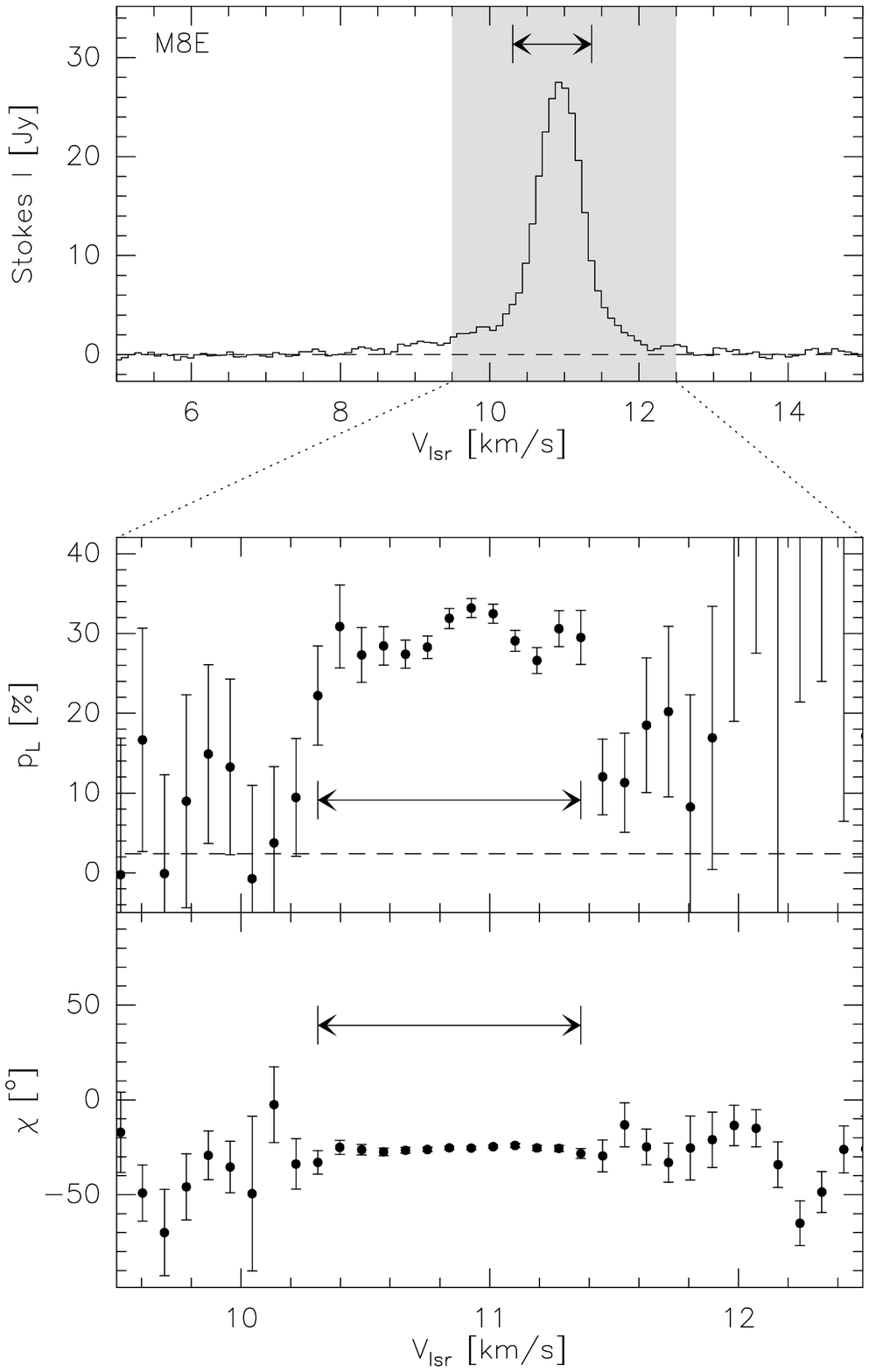}}
     \resizebox{8.5cm}{!}{\includegraphics*[bb=70 120 470 730]{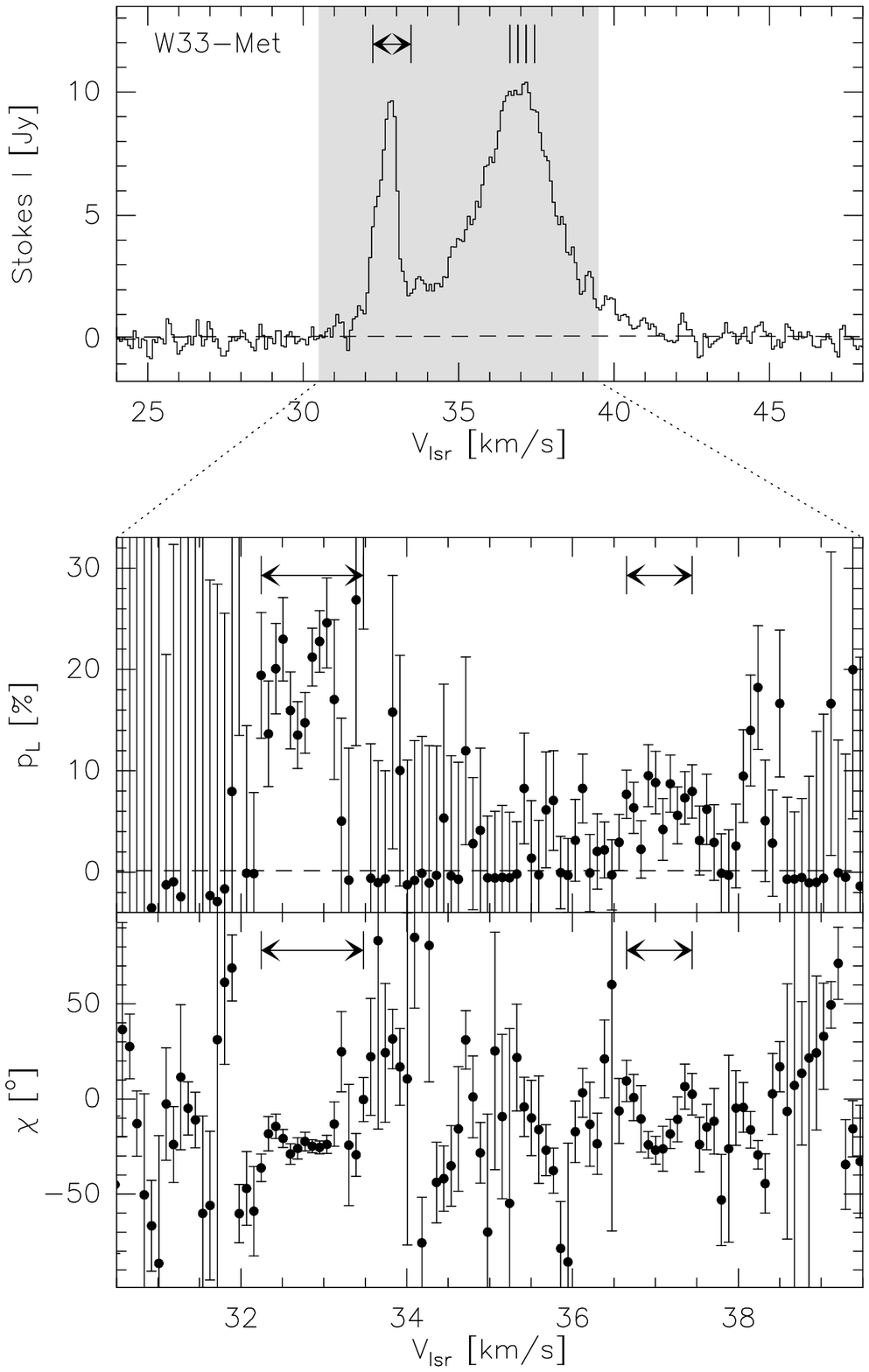}}
     \caption{Class~I methanol masers at 132.9\,GHz.\label{spectra_133ghz}}
   \end{figure*}
   \addtocounter{figure}{-1}
   \begin{figure*}
   \centering
     \resizebox{8.5cm}{!}{\includegraphics[bb=70 120 470 730]{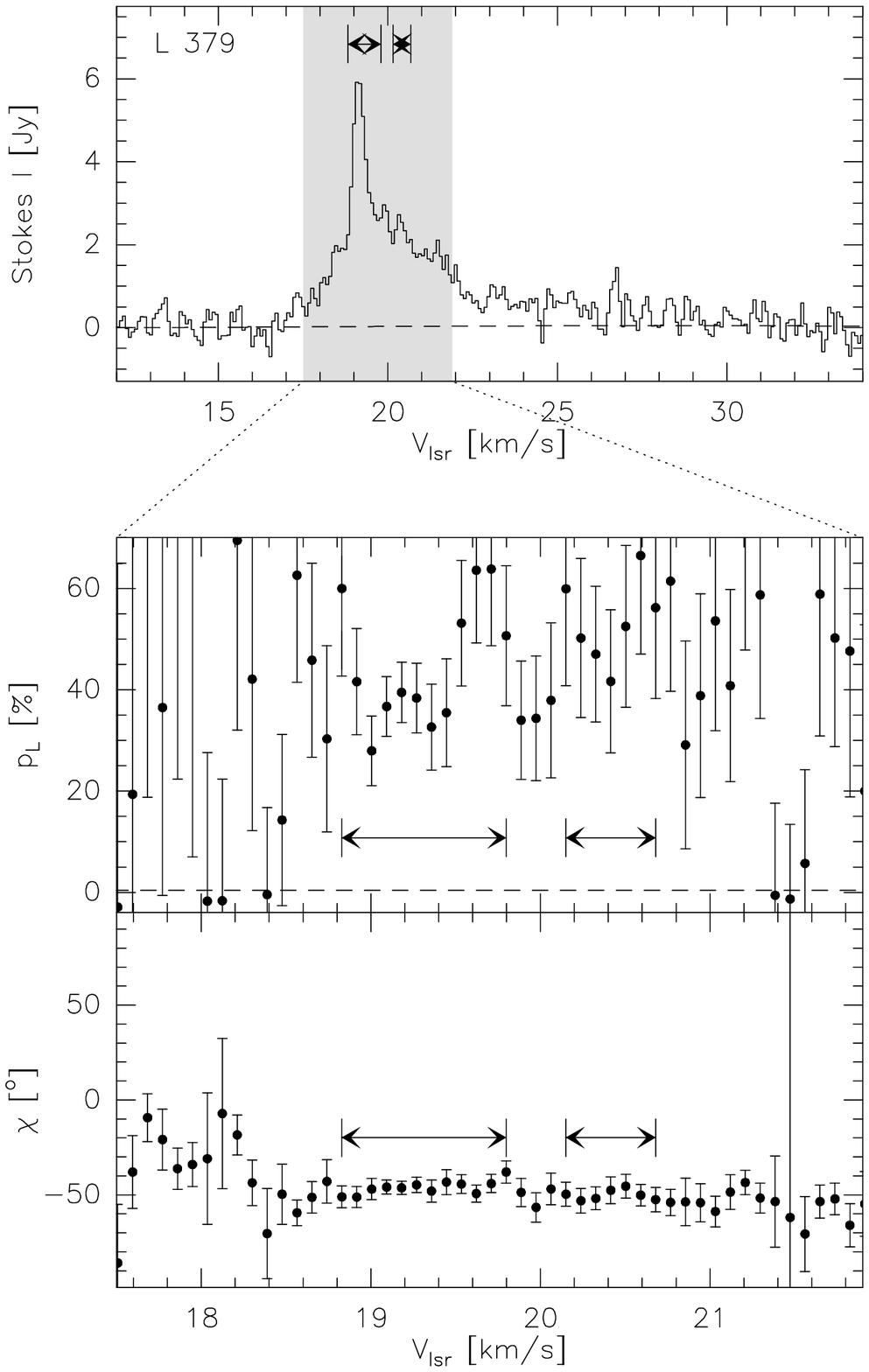}}
     \resizebox{8.5cm}{!}{\includegraphics[bb=70 120 470 730]{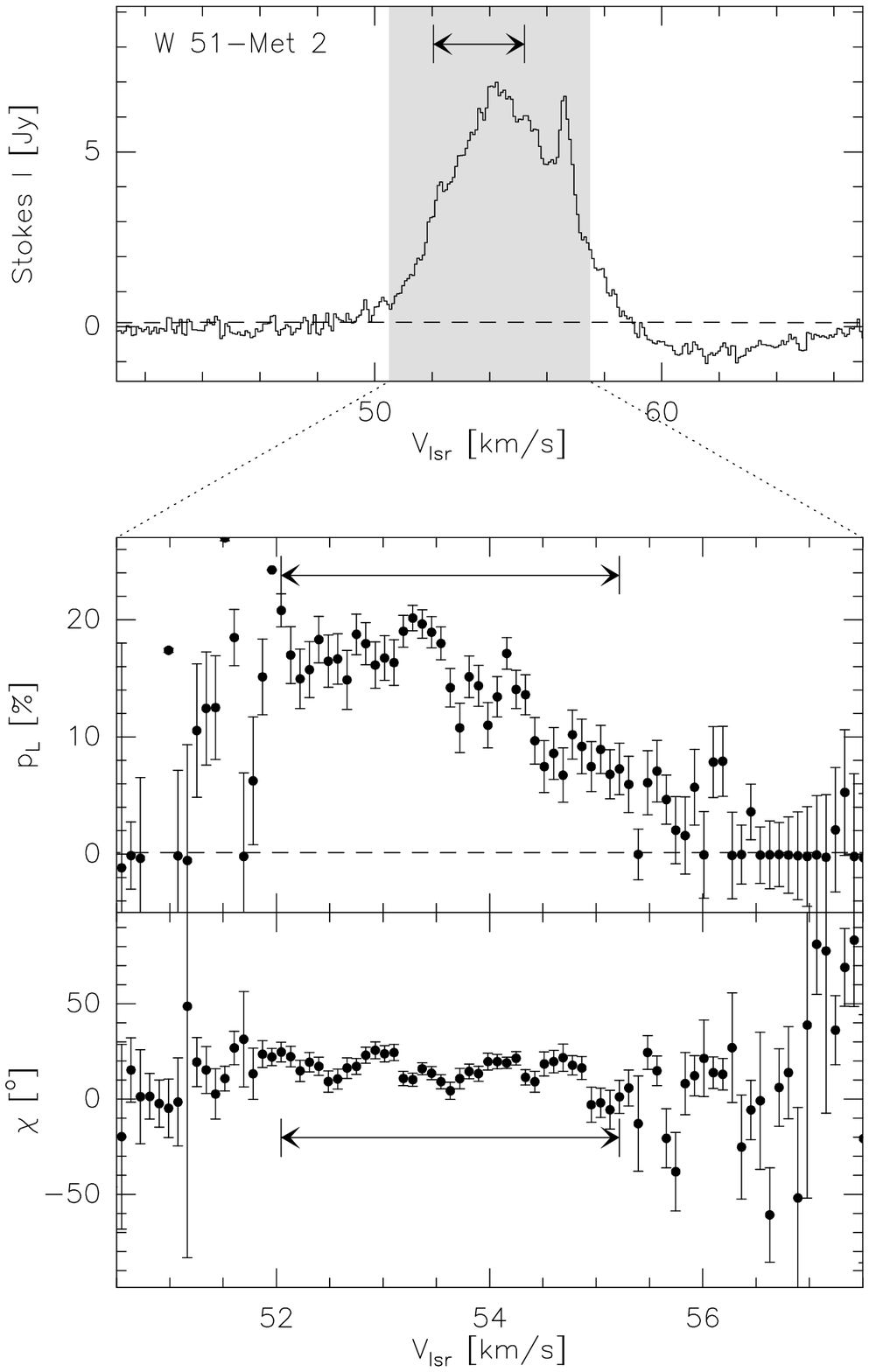}}
     \resizebox{8.5cm}{!}{\includegraphics[bb=70 120 470 730]{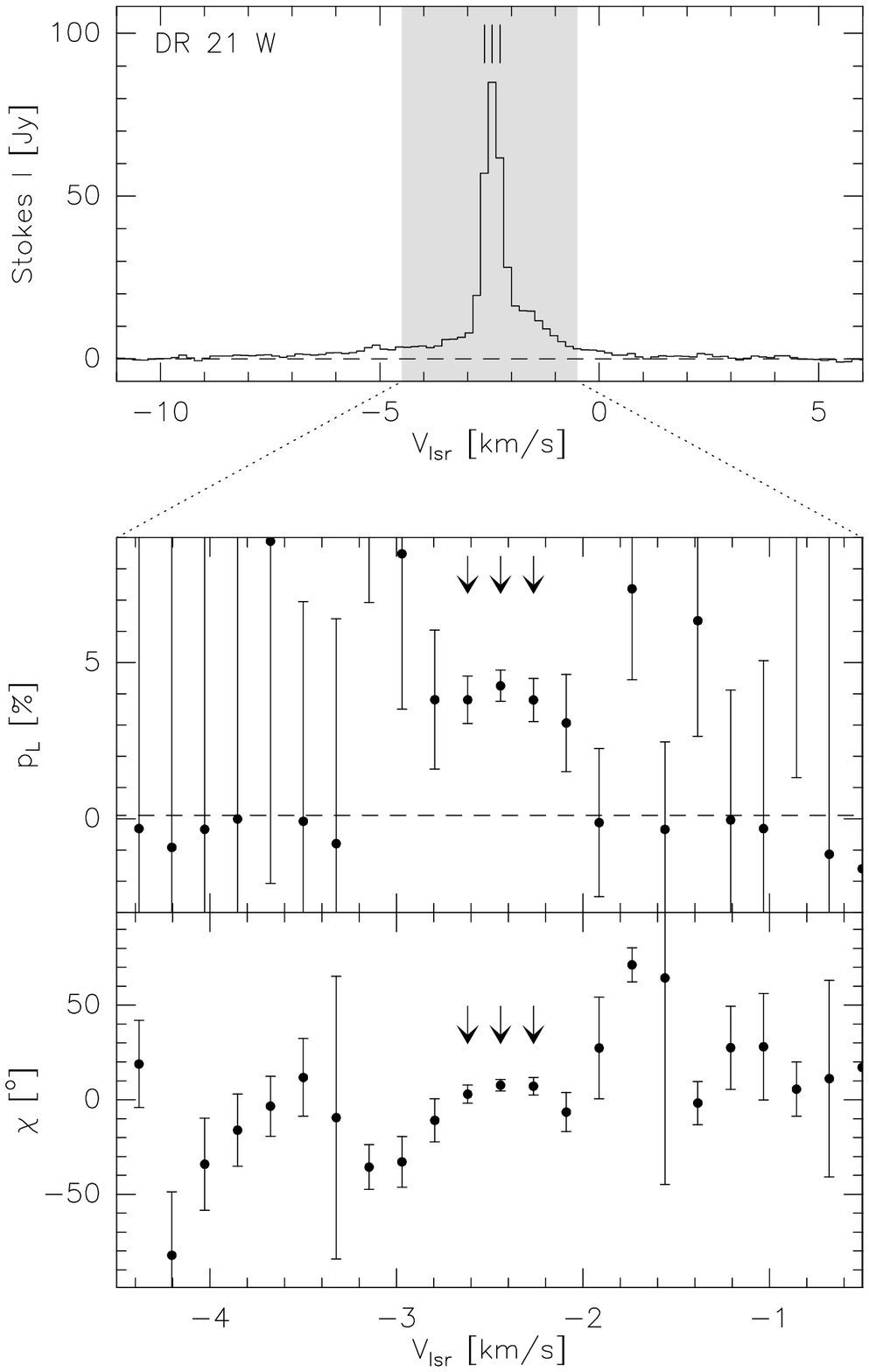}}
     \resizebox{8.5cm}{!}{\includegraphics[bb=70 120 470 730]{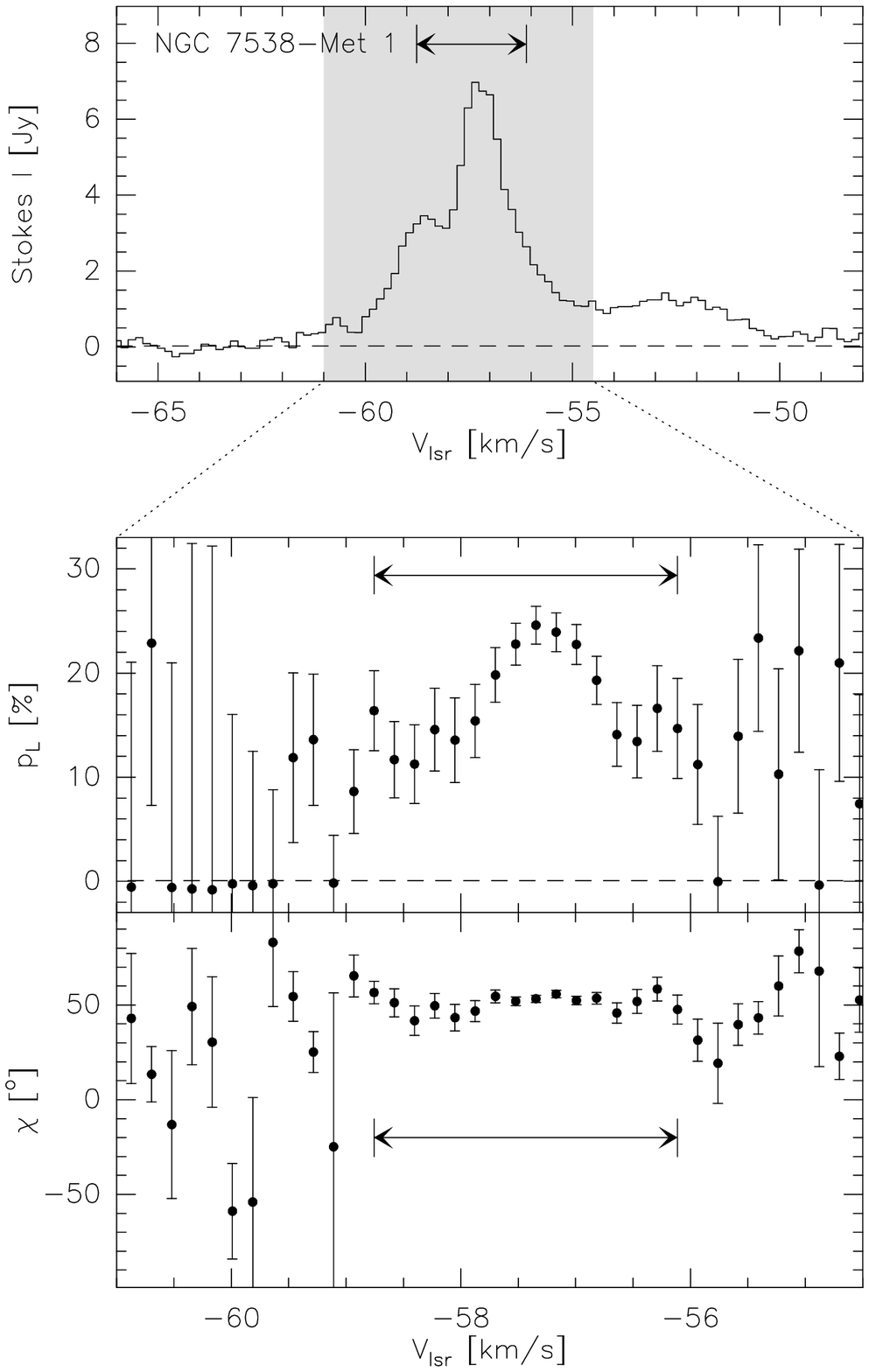}}
     \caption{(continued)}
   \end{figure*}

\subsubsection{\object{S\,231}} 
The 95\,GHz $8_0\rightarrow 7_1\,A^+$ \ch3oh  emission, first 
observed by Val'tts (\cite{1995ARep...39...18V}), is at $-16.8$\,\kms, with a 
3.8\,\% linear polarization.  Its polarization angle across the 
main component shows a clearly defined gradient, a finding that will be 
discussed in the next section. A second, weaker maser component, not or weakly
polarized, is at $-15.2$\,\kms, at the blue-shifted end of the cluster of 
6.7\,GHz $5_1\rightarrow 6_0\,A^+$ \ch3oh  masers (EVN observations of Minier 
et al., \cite{2000A&A...362.1093M}). 

\subsubsection{\object{NGC\,2264}} 
The 133\,GHz $6_{-1}-5_0\,E$ \ch3oh maser, first observed by Slysh et al.
(\cite{1997ApJ...478L..37S}), is located in a core of this massive star-forming
region, that is associated with the bright infrared source IRS\,1 (Allen, 
\cite{1972ApJ...172L..55A}). The position angle of the linear polarization 
($\sim 110^\circ$) remains remarkably constant. Schreyer et al. 
(\cite{1997A&A...326..347S}) decomposed the outflow activity of IRS\,1 
into two flows. They infer that one of them, extending in North-South 
direction, is driven by IRS\,1 and seen nearly pole-on. The 133\,GHz 
\ch3oh  maser is located in that outflow.

\subsubsection{\object{M8E}}
We find in the 133\,GHz line linear polarization as high as 33\,\%. Its 
position angle of $155^\circ$ is flat across the spectral profile. \object{M8E} 
harbours a strong 44\,GHz maser (Kogan \& Slysh (\cite{1998ApJ...497..800K}), 
whose two components are $1\farcs 7$ apart (2500\,AU at a distance of 1.5\,kpc, 
see Val'tts, \cite{1999ARep...43..148V}). The 133\,GHz maser is at the radial 
velocity of the strongest 44\,GHz component ($v_{\rm lsr} = 10.94$\,\kms,
the weaker one at $v_{\rm lsr} = 10.45$\,\kms does not appear at 133\,GHz). 
At 10.84\,\kms, we find a circular polarization of 
$p_{\rm C} = (-7.1 \pm 1.0)\,\%$, which cannot be due to the sidelobe 
polarization alone (Fig.~\ref{spectra_pc}). The maser is most likely not 
associated with the CO outflow (position angle $54^\circ$, W from N, Mitchell 
et al., \cite{1992ApJ...386..604M}), since the velocities of the quiescent gas 
and the methanol maser are close to each other (since the outflow almost 
appears pole-on, a large velocity offset would be expected for an associated 
maser).  Consequently, Val'tts (\cite{1999ARep...43..148V}) suggests that 
the Class~I maser is located in the front of the bipolar outflow at an 
interface with the surrounding dense molecular gas.

\subsubsection{\object{W\,33-Met}}
We measure, for this 133\,GHz maser, a fractional linear polarization of 
23\,\% at 32.9\,\kms. The radial velocity coincides with that of the southern
component of the two VLA 44\,GHz maser spots of Kogan \& Slysh 
(\cite{1998ApJ...497..800K}). Another polarized feature, with 
$p_{\rm L} \sim 10\,\%$, is at a radial velocity of about 37\,\kms, where the 
44\,GHz maser has a second component, at about $32''$ northwards. This feature 
coincides with a broader \ch3oh  emission, of equal flux density. 
Val'tts (\cite{1999ARep...43..157V}) suggests that this (spatially extended) 
component is a quasi-maser with a low gain and weak line narrowing. Thus, the 
linear polarization measured by us (although barely significant, 
$\sim 3\sigma_{\rm rms}$) towards this component almost certainly has an 
important instrumental contribution. 

No spatially resolved map of the outflow exists. Val'tts 
(\cite{1999ARep...43..157V}) find the Class~I maser to be associated with
a CS condensation, containing a very young protostar.

\subsubsection{\object{L\,379}}
The 133\,GHz maser is located $30"$ to the north of IRAS source 
\object{L\,379-IRS1}. From the absence of linear polarization in the thermal 
red-shifted linewing, it is excluded that the measured linear polarization of 
39\,\% at a radial velocity of 19.2\,\kms is due to the telescope's sidelobe 
polarization. 
The polarization angle has a flat spectral profile with $\chi = 134^{\circ}$.
This 133\,GHz maser concides with the northern $\lambda\,800\,\mu$m continuum 
peak, discovered by Kelly \& MacDonald (\cite{1996MNRAS.282..401K}), and is 
probably associated with their CO outflow, whose direction is at $75^\circ$.  
It is likely that the red-shifted linewing consists of a blend of 
(quasi-)maser emission. As a matter of fact, we find an even stronger 
fractional linear polarization with a similar polarization angle at about 
19.6\,\kms, but of lower significance ($3-4\,\sigma_{\rm rms}$). 
The 44\,GHz \ch3oh  emission (Kogan \& Slysh, \cite{1998ApJ...497..800K}) shows 
a total of 11 masers, scattered across a region of $46''$ in diameter, in two 
groups at radial velocities of $18.8-20.8$\,\kms and $16.3-18.8$\,\kms, 
respectively. From the latter group, we detect \ch3oh  emission, whose line 
profile can be decomposed into two Gaussian spectral components, a large one 
(linewidth 7.0\,\kms {\sc fwhm}), and a narrower one, with a linewidth of 
2.1\,\kms (at best a quasi-maser without substantial line narrowing). The 
polarization of these components, most likely scattered across our beam, is 
strong across the whole line profile. Stokes U is an inverted and scaled copy 
of Stokes I, and the signs of the other Stokes parameters agree with the 
instrumental signature. We thus discard this polarization signal as of 
instrumental origin.

\subsubsection{\object{W\,51-Met\,2}}
The 84\,GHz line of this \ch3oh  maser is blended with a broad thermal 
emission. However, the linear polarization disappears across a large part of 
the thermal profile, except for the single-peaked maser line at 56.1\,\kms, and
another component at 53.3\,\kms (possible maser emission masked by the thermal 
emission, cf. Fig.~\ref{spectra_84ghz}). The spectral profile of the 
polarization angle of the 84\,GHz maser spike has a linear slope, as in the case
of \object{S\,231}. Jumps in the polarization angle of $90^\circ$ at certain 
velocities are due to a statistical bias in the weak-signal limit. The 
thermal emission of the 133\,GHz maser is strongly polarized (between 52.0 and 
55.2\,\kms, reaching a maximum of 20.1\,\% at 53.3\,\kms, with a polarization 
angle of $-5^\circ$ to $30^\circ$ in this velocity range). The line peak can be 
decomposed into a broad component (linewidth 4.5\,\kms {\sc fwhm}) and a narrow
one (linewidth 0.482\,\kms). We consider it quite possible that this 
extraordinary, broad polarization profile may be due to an anisotropic optical 
depth (Goldreich \& Kylafis, \cite{1982ApJ...253..606G}). The amplification of 
polarized background continuum serving as "seed" is also 
possible, given the flux density ratio of $F_{\rm line}/F_{\rm cont} = 11$ at 
53.3\,\kms (or 2.8 if we only consider the narrow component).  
However, the uncertain astrometry and extent of the thermal emission could 
introduce an important instrumental contribution, too (as discussed in 
the previous section), which makes impossible an unbiased detection of the 
polarization of the 133\,GHz maser at 56.6\,\kms (the linear polarization 
is not significant, as opposed to the 84\,GHz maser).

As an indication of the possibility of polarized seed radiation, we detect a 
fractional linear continuum polarization of $p_{\rm L} = (4.7 \pm 1.2)\,\%$ at 
84\,GHz, and $p_{\rm L} = (13.7 \pm 2.1)\,\%$ at 133\,GHz (with 
$\chi = 18\fdg 4$, not too far from the submm measurement, Chrysostomou et al.,
\cite{2002A&A...385.1014C}), respectively. Note that the errors in $p_{\rm L}$ 
do not contain the uncertainty due to atmospheric fluctuations (as discussed in
Section\,2).

The host source is the \object{W\,51\,f} subregion of the \object{W\,51\,A} 
star forming cloud (see Sievers et al., \cite{1991A&A...251..231S}, for a 
$\lambda\,870\,\mu$m and $\lambda\,1.3$\,mm map). Submillimeter continuum 
polarimetry (Chrysostomou et al., \cite{2002A&A...385.1014C}) reveals a 
marginal linear dust polarization of 2.5\,\%, at a position angle $24^\circ$. 
The authors conclude that the overall magnetic field is rather weak, and 
becomes only important close to the sites of star formation. The lack of 
evidence for a continuum signal at longer wavelengths made it difficult to 
reveal the nature of this maser's environment. Hodapp \& Davis 
(\cite{2002ApJ...575..291H}) recently found a spatially unresolved, but strong 
H$_2\,(1-0)$ S(1) line ($\lambda\,2.12\,\mu$m) about $10"$ eastwards of the 
methanol maser position. Taking into account the uncertain astrometry (no 
interferometer observations are available), this emission could come from the 
same shocked gas that harbours the methanol maser, in the immediate vicinity of 
a star enshrouded by the molecular and dust cloud. 

\subsubsection{\object{DR\,21-W}}
\object{DR\,21-W} is one of three sources in which we heave measured and 
detected linear polarization in both the 84\,GHz and 133\,GHz line. They both 
have flat polarization angle profiles (with position angles $48\fdg 1$ and 
$7\fdg 7$ respectively). They are single-peaked, and are rather weakly 
polarized (about $3-4$\,\%). At 133\,GHz, we find a significant circular 
polarization of $p_{\rm C} = (3.52 \pm 0.68)\,\%$ at 
$v_{\rm lsr} = -2.63$\,\kms (Fig.~\ref{spectra_pc}), slightly offset from the 
velocity of maximal linear polarization ($-2.44$\,\kms). As in the case of 
\object{M8E}, this detection can not be entirely attributed to the sidelobe 
polarization.

Kogan \& Slysh (\cite{1998ApJ...497..800K}) observed the $7_0-6_1\,A^+$ 
methanol maser at 44\,GHz with the VLA, and find two major components separated
by $2\farcs 8$, whose radial velocities are close to ours. The environment of 
this maser is fairly well known: Davis \& Smith (\cite{1996A&A...310..961D}) 
obtained high-resolution images in H$_2\,(2.12\,\mu$m emission).
Their spectrum closest to the methanol maser is best described by a C-type 
planar shock. The submillimetre continuum polarimetry (Minchin \& Murray,
\cite{1994A&A...286..579M}) reveals a linear magnetic field, in position angle 
($114^\circ$) intermediate between the cloud core and the dominant outflow axis
($70^\circ$, Garden et al., \cite{1991ApJ...366..474G}). This polarization 
angle, however, does not agree well with that of either of the two maser lines.
Liechti \& Walmsley (\cite{1997A&A...321..625L}) observed the thermal 
$2_{\rm k} \rightarrow 1_{\rm k}$ emission of $A^+$ and $E$-type methanol. The 
methanol maser shown here coincides in position with a thermal clump in 
$2_1 \rightarrow 1_1\,E$ emission, whose major axis is at position angle 
$45^\circ$ - remarkably close to our 84\,GHz polarization angle. 
The thermal methanol emission also appears in our 84\,GHz and 133\,GHz spectra
as a broad, unpolarized  pedestal underneath the maser lines. 

\subsubsection{\object{W\,75 S(3)}}
In this source, located in the same ridge of molecular gas as the 
\object{DR\,21} region (Wilson \& Mauersberger, \cite{1990A&A...239..305W}), 
the 84\,GHz masers consist of a narrow main component, and a blend of several, 
weaker (quasi-) maser components, together with a broad underlying thermal 
emission. Only the polarization of one of the weaker components is significant,
the other maxima of the polarization profile are far less significant (see 
Fig.~\ref{spectra_84ghz}). It is difficult to distinguish this feature from an 
intrinsic or instrumental polarization of the underlying, broad emission. The 
profile of the polarization angle oscillates between $50^\circ$ and $80^\circ$. 
At 133\,GHz, no significant linear polarization was detected.

\subsubsection{\object{NGC\,7538-Met1}}
This is the only source in our sample with a bona-fide detection of maser 
polarization in each of the three Class~I maser frequencies. The polarization 
angles at the maser peaks are different, their velocities are within 0.8\,\kms.
This difference is most likely intrinsic, since the 
polarization angle profiles themselves have a peak-to-peak variation of up to 
$20^\circ$. The strongest polarization is found at 133\,GHz (24.6\,\%),
resulting in the most significant polarization angle measurement among the 
three transitions. The polarization angle has a remarkably flat spectral profile
(position angle $53\fdg 3$ at the 133\,GHz maser spike, as compared to 
$161\fdg 2$ at 84\,GHz, and $7\fdg 2$ at 95\,GHz). Although part of the 
polarization could come from the underlying broad emission (at 84\,GHz and 
133\,GHz), it is unlikely that the latter reaches a fractional polarization as 
high as 25\,\%. The $\lambda\,870\,\mu$m continuum polarization is from 
magnetically aligned dust grains (Momose et al., \cite{2001ApJ...555..855M}). 
Its position angle ($41\fdg 8$) differs by $11\fdg 5$ from our measurement at 
133\,GHz (Figure~\ref{ngc7538}). 

\subsection{Class~II Masers}
\begin{figure}
     \resizebox{8.5cm}{!}{\includegraphics[bb=70 120 470 730]{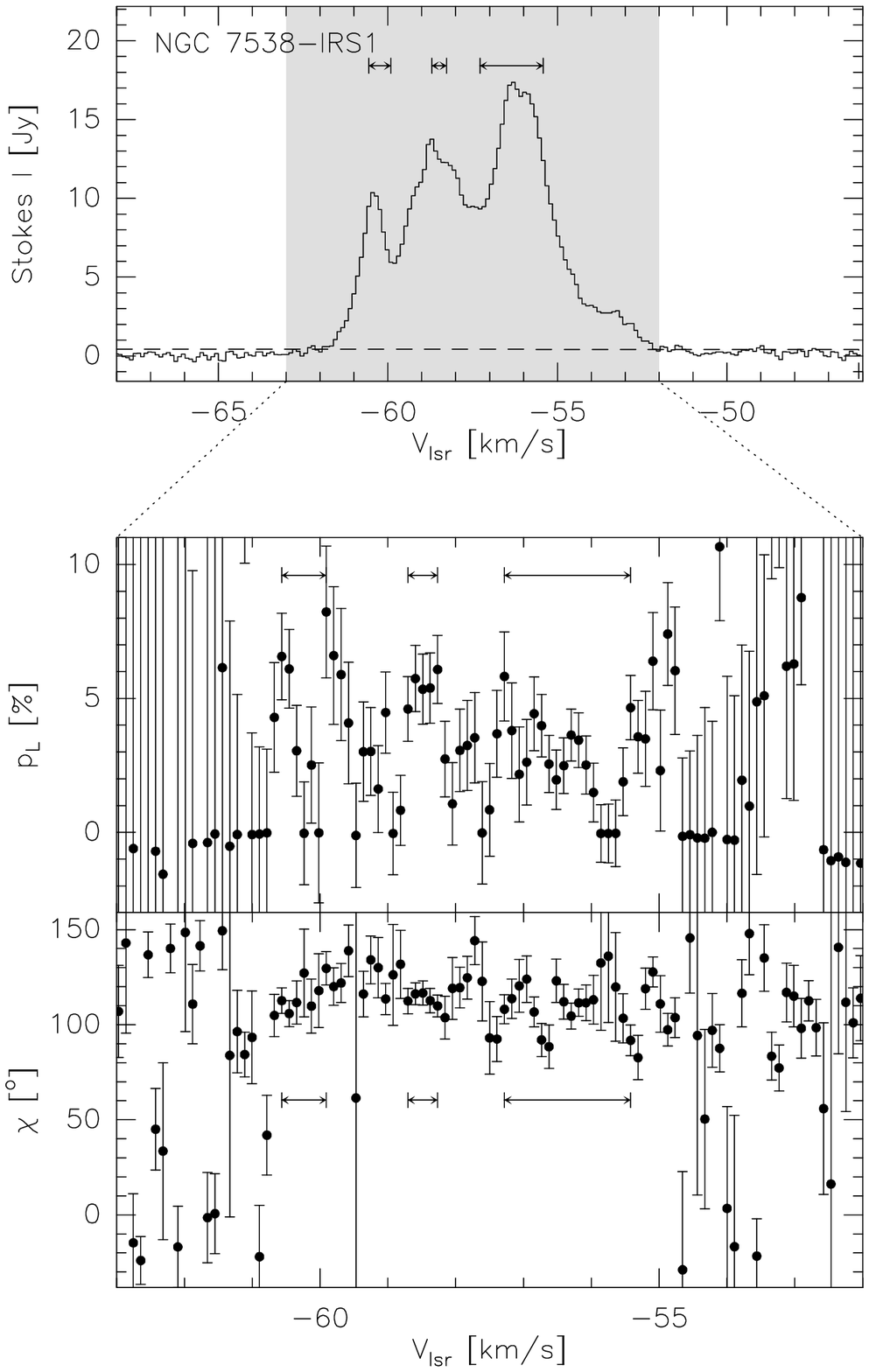}}
     \caption{The 107~GHz polarization spectra of \object{NGC\,7538-IRS1}, a 
     Class~II \ch3oh  maser. \label{spectra_107ghz}}
   \end{figure}

\begin{figure*}
   \centering
     \resizebox{8.5cm}{!}{\includegraphics[bb=70 120 470 730]{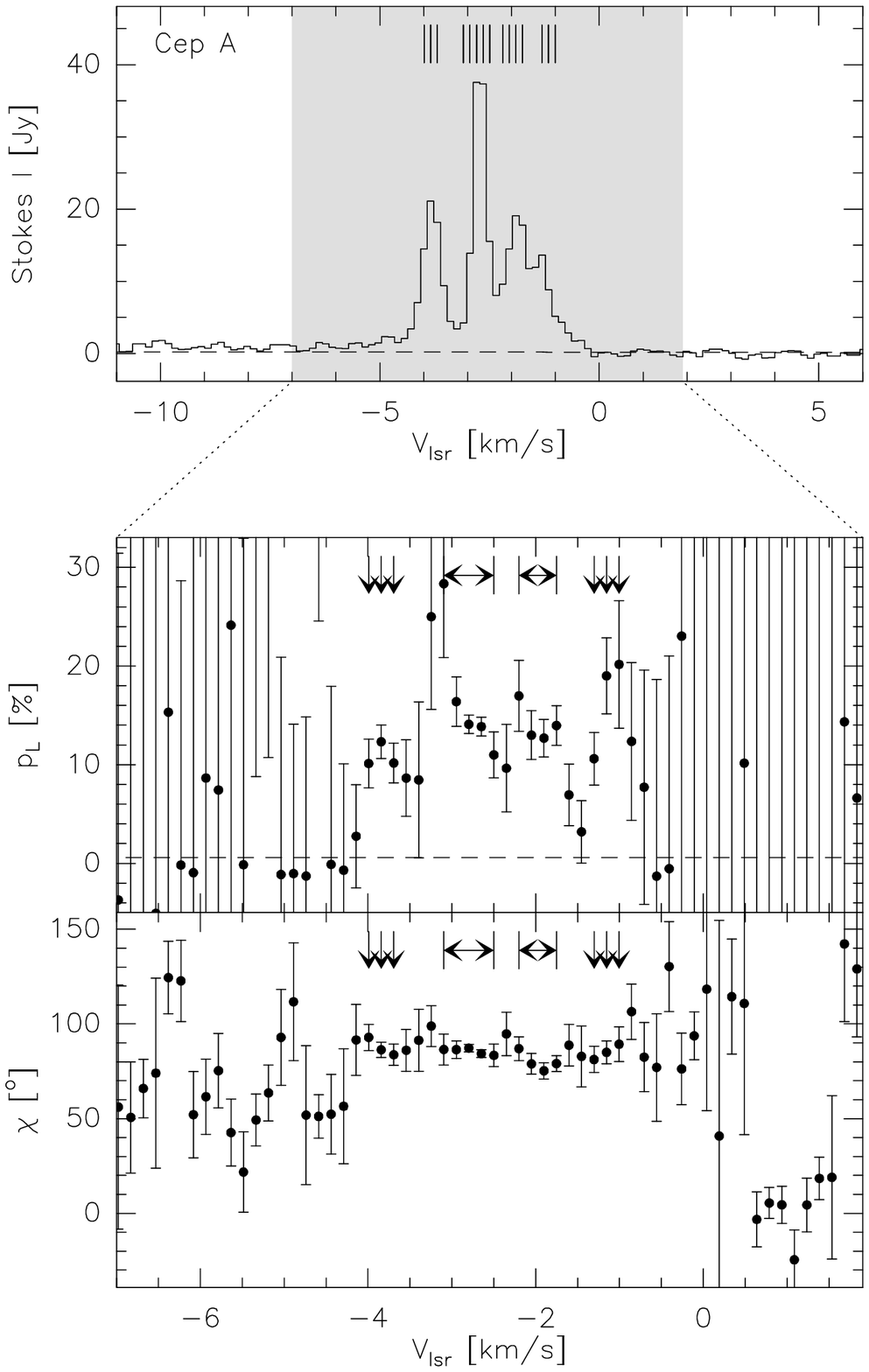}}
     \resizebox{8.5cm}{!}{\includegraphics[bb=70 120 470 730]{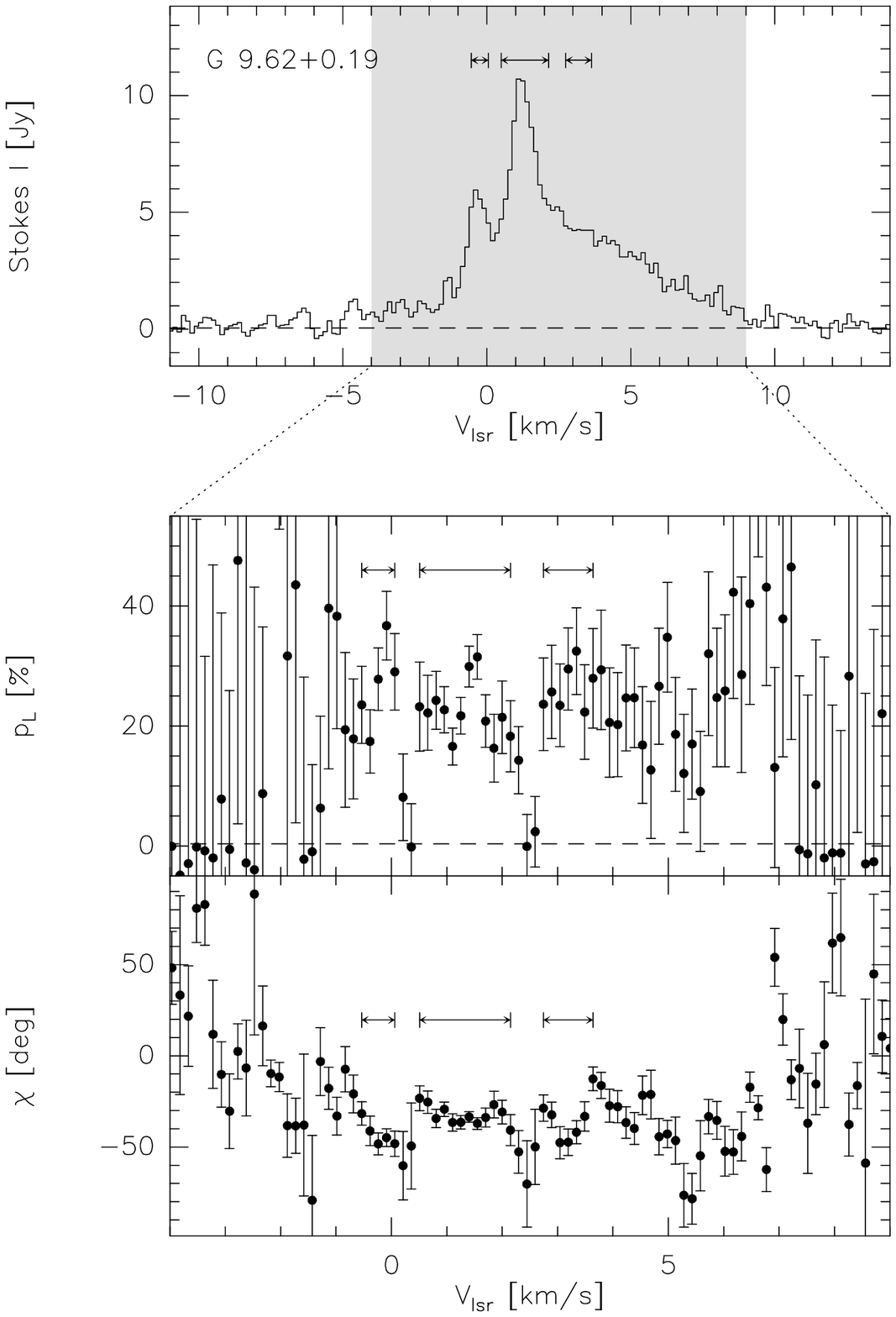}}
     \caption{Class~II methanol masers at 157.0\,GHz. 
     \label{spectra_157ghz}}
   \end{figure*}

The polarization spectra for Class~II masers are shown in
Figures\,\ref{spectra_107ghz} and \ref{spectra_157ghz}, for 107\,GHz
and 157\,GHz, respectively.
\subsubsection{\object{NGC\,7538-IRS1}}
The 107\,GHz maser from this source is located $37''$ northwards of its 
Class~I counterpart, and coincides with the infrared source IRS\,1. The 
107\,GHz methanol maser has three components, each of similar polarization. The 107\,GHz continuum of the source, as seen in our spectra, is not polarized
(upper limit $p_{\rm L} < 0.4\,\%$). The velocities of the maser spikes
correspond to those of the 6.7\,GHz maser components A,E, and D (or C) of 
Minier et al. (\cite{2000A&A...362.1093M}), which are $0\farcs 6$ apart (i.e. 
1620\,AU at 2.7\,kpc distance). The polarization angle of the three velocity 
components of the 107\,GHz maser (Figure~\ref{spectra_107ghz})
is $105^\circ -116^\circ $, not too far from the 
position angle of the aligned 6.7\,GHz masers of subgroup A, and close to the 
position angle of the equator of the edge-on circumstellar disk proposed by 
Minier et al. They suggest that the maser resides at a single radius of this 
disk. This conjecture is corroborated by the detection of an elongated 
structure seen in an interferometric $^{13}\,\mathrm{CO}\,(1-0)$ map (Scoville 
et al., \cite{1986ApJ...303..416S}), perpendicular to the bipolar H\,II 
region, and at an angle of $55^\circ$ to the major outflow axis. As the 
Class~I maser to the south of IRS\,1, our Class~II maser polarization angle 
agrees more or less (within $10^\circ$ to $20^\circ$) with that of the 
$\lambda\,870\,\mu$m dust emission (Momose et al., \cite{2001ApJ...555..855M}, 
see our Figure~\ref{ngc7538}). The linear polarization at 107\,GHz yields
significant results for three velocity components ($p_{\rm L} = 3.6$ to 
6.1\,\%, Fig.~\ref{spectra_107ghz}). However, given that this fractional 
polarization is rather weak (for the 157\,GHZ maser, no significant 
polarization is detected), it cannot be excluded that the whole line profile, 
with both (quasi-) maser and thermal emission is polarized, due to the 
amplification of a polarized continuum source. As a matter of fact, the 
continuum emission at 157\,GHz has a linear polarization of 
$p_{\rm L} = (8.4 \pm 1.0)$\,\%. Its position angle ($-84\fdg 0 \pm 2\fdg 7$) 
is close to that of the $\lambda\,870\,\mu$m polarization ($-87\fdg 2$, Momose 
et al.), which differs from the polarization angle of the 107\,GHz maser by 
about $23^\circ$. For the ratio of the line to continuum brightness 
temperature we estimate values of 6 for the 107\,GHz maser, and of 5 for the 
157\,GHz maser, respectively.

\subsubsection{\object{Cep\,A}}
We observed the 107\,GHz and 157\,GHz masers towards \object{Cep\,A East}, 
detecting a significant linear polarization only from the latter. This source 
contains masers at four well separated velocities, which agree with the 
components of the 12.2\,GHz maser (VLBI observation of Minier et al., 
\cite{2000A&A...362.1093M}). The fractional polarization of the components of 
the 157\,GHz masers ranges from $p_{\rm L} = 10.6$\,\% to 14.1\,\%, and the 
polarization angle from $79^\circ$ to $86^\circ$ (see Fig.\,\ref{cepa_zoom}). 
The significant polarization measurements (i.e. $p_{\rm L} > 3\sigma$) are 
clearly correlated with the maser spikes (there is stronger, but less 
significant polarization between the spikes). The velocity of the first 
component listed in Table\,\ref{maser_survey_class2} agrees well with that of 
component B in the interferometer map of the 107\,GHz maser (Mehringer et al.,
\cite{1997ApJ...475L..57M}). Our last two components correspond in both 
velocity and flux density to component A of Mehringer et al. (after smoothing 
to their spectral charactersistics), which is spectrally unresolved by them, but
asymmetric due to the blend of the two spikes. Although Mehringer et al. infer 
from their results that the 107\,GHz masers arise from regions with slightly 
different physical characteristics, their polarization properties are rather 
homogeneous - although the projected distance between components A and B is 
$0\farcs 7$ (corresponding to 500\,AU at a distance of 730\,pc, 
Torrelles et al., \cite{1996ApJ...457L.107T}). The fourth component at 
$v_{\rm lsr} = -2.8$\,\kms   does not have a counterpart at 107\,GHz. At 
157\,GHz, it is the strongest one (both in intensity and polarization).   
The velocities of all these masers are redshifted with respect to those
of the ambient molecular cloud and the expanding bubble of H$_2$O masers 
(Torrelles et al., \cite{2001Natur.411..277T}). According to Minier et al. 
(\cite{2000A&A...362.1093M}), the methanol masers and the water maser are 
located in the same environment. However, given the complexity of the region, 
it cannot be ruled out that the \ch3oh  maser rather emerges close to the  
thermal radio jet of Torrelles et al. (\cite{1996ApJ...457L.107T}). Codella 
et al. (\cite{2003MNRAS.341..707C}) found evidence of a well collimated outflow
towards the south; the methanol maser could be located just at its onset.  

\subsubsection{\object{G\,9.62+0.19}}
This maser has two components on top of a broad thermal and possibly 
quasi-maser contribution. As in the case of \object{Cep\,A} and 
\object{NGC\,7538-IRS1}, both velocity components of this maser have the same 
polarization characteristics (weighted mean $p_{\rm L} = 33.0$\,\%, at 
$\chi = -140\fdg 6$). The masers' position and velocity coincide with that of a 
cluster ($0\farcs 5$ in extent) of 6.7\,GHz maser emission (Norris et al., 
\cite{1993ApJ...412..222N}). The associated UC\,H{\sc ii} region of this 
methanol maser is continuum source E (Garay et al., \cite{1993ApJ...418..368G}),
located near the edge of a compact, dense and warm molecular cloud core (Hofner
et al., \cite{1996ApJ...460..359H}). Hot thermal methanol emission of the 
$15_3\rightarrow 14_4\,A^-$ line was imaged with the Plateau de Bure 
interferometer by Hofner et al. (\cite{2001ApJ...549..425H}). It coincides with
the continuum source E and the maser position.
Although the thermal methanol emission is point-like in the interferometer map, 
there may be an extended component missed by the interferometer. As a matter 
of fact, the whole line profile at 157\,GHz is polarized, which may be due to 
the sidelobe polarization of a component that is resolved in our beam. 
 
\section{Discussion}
Most models of maser polarization to date have been for
saturated masers (Goldreich et al., \cite{1973ApJ...179..111G}, 
Watson \& Wyld, \cite{2001ApJ...558L..55W}). The polarization in such models 
is due to unequal populations in Zeeman--split substates of the maser levels 
and one finds that a high degree of maser saturation as well as the presence
of a magnetic field is a necessary condition for linear polarization in this 
type of model. One can also have polarization in other situations, where 
the radiation field pumping the maser or the optical depth in FIR 
transitions connected to one of the maser levels is anisotropic. 
Western \& Watson (\cite{1983ApJ...275..195W}, \cite{1984ApJ...285..158W})
investigated the role of anisotropic pumping in the polarization of maser 
lines, both for $J=1-0$ and $J=2-1$ transitions. Elitzur 
(\cite{1993ApJ...416..256E}) considered the exponential maser growth in the 
unsaturated phase, where the polarization is produced. 
Subsequently, it has been found (Elitzur, \cite{1996ApJ...457..415E}) that for 
non-paramagnetic molecules, where linewidths exceed the Zeeman 
splitting, the higher-order terms of the Zeeman splitting, although small, 
cannot be neglected, but have to be taken into account in order to rule out 
unphysical solutions. A summary of this work can be found in Elitzur 
(\cite{2002apsp.conf..225E}).

In the case of the masers observed by us, we feel that one should presently 
consider all options open and concentrate on obtaining better observational
constraints on the physical conditions in the regions where methanol maser 
polarization is observed. Below we comment briefly both on what is 
known about the physical conditions in the clumps giving rise to methanol 
masers as well as on some of the current models.

\subsection{Physical conditions in the regions giving rise to methanol maser 
emission}
Studies of Class~I methanol masers to date have been hampered by poorly known 
collisional rates. However, a recent study by Pottage, Flower \& Davis 
(\cite{2002JPhB...35.2541P}, \cite{2004JPhB...37..165P}) of He-\ch3oh  
collisions has improved the situation, and it appears as if the estimates 
used in previous studies (e.g. Cragg et al., \cite{1992MNRAS.259..203C},
Johnston et al., \cite{1992ApJ...385..232J}) were broadly correct. Calculations
by Leurini et al. (\cite{2004A&A...422..573L}) using the new rates confirm 
this.

On the basis of the earlier rate estimates, Johnston et al. found, using VLA 
observations of the 25 GHz transitions towards Orion, that the H$_2$ density in 
the maser region should be between  $2\times 10^{6}$ and $10^{8}$ cm$^{-3}$ 
and the methanol column density in the range $2\times 10^{16}$ to 
$2\times 10^{17}$ cm$^{-2}$. The temperature is expected to be of order 100~K 
and the inferred methanol abundance \ch3oh/H$_2$ $10^{-7}$ to $10^{-6}$. These 
results are similar to those obtained for other \ch3oh  lines in other Class~I
sources (e.g. Liechti \& Walmsley, \cite{1997A&A...321..625L}, in the case of 
DR21) and one concludes that high density (above $10^5$ cm$^{-3}$), high 
temperature (of order 100~K), and high methanol abundance are characteristics 
of Class~I methanol emission. It is notable moreover that the estimated 
abundances are similar to those found by Bachiller et al. 
(\cite{1995A&A...295L..51B}) towards outflow regions, but much higher than those
in cold dark clouds (Friberg et al., \cite{1988A&A...195..281F}). One 
interpretation of this is that the elevated methanol abundance is caused by 
sputtering of ice mantles of dust grains containing methanol in shocks 
associated with the outflows. If so, the alignment due to the shock likely has 
consequences for the magnetic field orientation and for the escape probability 
in optically thick \ch3oh  lines.

Class~II methanol masers (in particular the 6.7 and 12.2 GHz transitions) have 
been studied by Sobolev et al. (\cite{1997MNRAS.288L..39S}, 
\cite{1997A&A...324..211S}) and by Cragg, Sobolev \& Godfrey 
(\cite{2002MNRAS.331..521C}). The general conclusion is that the masers form in
high density regions ($10^5$ to $10^8$ cm$^{-3}$) under conditions in which the
gas temperature is lower than the dust temperature (both being of order 100~K).
There is evidence for association with OH maser regions having milligauss 
magnetic fields. There is evidence also that in W3OH, the maser regions are 
associated with absorption components in non--masing transitions of OH 
(Guilloteau et al., \cite{1985A&A...153..179G}) and methanol (Menten et al., 
\cite{1986A&A...169..271M}) whose temperatures are of the order $100-200$~K. In
both cases (\ch3oh  and OH), one requires high abundances in the range 
$10^{-6}$ to $10^{-7}$ whose explanation (Hartquist et al., 
\cite{1995MNRAS.272..184H}) is thought to lie in the evaporation of \ch3oh  
from the surface of grains heated by the passage of weak shocks. As a matter of
fact, ISO observations (Gibb et al., \cite{2000ApJ...536..347G}) have shown 
that \ch3oh  is an important constituent of interstellar ices associated with 
protostellar objects such as \object{W\,33\,A}. For low mass stars, Pontoppidan
et al.  (\cite{2003A&A...404L..17P}) detect abundant solid methanol, too.  

\subsection{Models of polarization of saturated masers}
Goldreich et al. (\cite{1973ApJ...179..111G}) showed that the Zeeman splitting 
$\Omega_{\rm B}$ in polarized masers has to exceed $\sqrt{R\Gamma}$, where
$R$ is the stimulated emission rate of the masing transition, and $\Gamma$ is 
the decay rate out of the maser levels. However, this result only holds for 
$J=1-0$ transitions. For higher angular momenta, Deguchi \& Watson 
(\cite{1990ApJ...354..649D}) showed that this condition has to be replaced by 
the stronger requirement $\Omega_{\rm B} > R$. It is difficult to get 
even a rough estimate of $R$ from observable quantities, since the spot size 
of mm \ch3oh  masers is not well known. Lonsdale et al. 
(\cite{1998AAS...193.7101L}) deduce from mm VLBI measurements a lower limit of 
several milliarcseconds. In the following, we assume that $R \la \Gamma$, as 
for unsaturated maser emission. We estimate for the gas 
density in \ch3oh  masers values of $10^6-10^7\,{\rm cm}^{-3}$, where magnetic 
fields of around a 1\,mG are expected. The Zeeman splitting is thus about 
0.3\,Hz. For the Land\'e factor of \ch3oh, a value of 0.078 has been used 
(Jen, \cite{1951PhRv...81..197J}), or equivalently $4.2\times 10^{-5}$ if the 
Bohr magneton is used instead of the nuclear magneton. 
The loss rate $\Gamma$, on the other hand, is around 
$10^{-4} - 10^{-3}\,{\rm s}^{-1}$, for a collision rate of 
$10^{-10}\,{\rm cm}^{-3}n_{\rm H_2}$, and the above gas density.  
The Zeeman splitting thus exceeds R by several orders of magnitude, 
except if the maser is strongly saturated (which is not the case here).

According to Goldreich et al. (\cite{1973ApJ...179..111G}), the polarization 
angle is either perpendicular to the projection of the magnetic field on the 
sky (hereafter ${\bf B}_{\rm sky}$), if $\sin^2{\Theta} > 2/3$, where 
$\Theta$ is the angle between the magnetic field direction and the maser 
propagation direction, or parallel otherwise. However, this result only holds 
in the limiting case of extreme saturation $\Gamma << R$ and magnetic fields
satisfying $R << \Omega_{\rm B}$.

Extensions of this model were studied by Nedoluha \& Watson 
(\cite{1990ApJ...354..660N}, hereafter NW90), including lower saturation 
and larger angular momenta. Since they are relevant for our results, 
we discuss them in the next section.

\subsection{Extensions of the idealizing maser polarization models}

NW90 (their Fig.\,3a,b) showed that for all saturation parameters $R/\Gamma$ 
and for a maser propagation with $\sin^2{\Theta} > 2/3$ (yielding the largest
linear polarization) the polarization angle remains (except for a small 
deflection) roughly perpendicular to ${\bf B}_{\rm sky}$. For smaller
$\Theta$, all orientations of the polarization angle, from parallel to 
${\bf B}_{\rm sky}$ for weak saturation to perpendicular to 
${\bf B}_{\rm sky}$ for strong saturation, are possible. For a $J=2-1$ 
transition, the polarization is lower than for a $J=1-0$ transition. We thus 
infer that the large polarizations ($\ga 10\,\%$) observed by us towards the 
Class~I masers would require $R \sim 100\,\Gamma$, intermediate magnetic fields
($\Omega_{\rm B}/\Gamma \sim 100$) and propagation preferentially 
perpendicularly to the magnetic field (cf.  Fig.\,3{\it c,d} of NW90). A 
stronger polarization can only be achieved by considering bidirectional 
propagation. According to NW90, for $J=2-1$ transitions, $R \sim \Gamma$, and 
propagation perpendicularly to the magnetic field, $\sim 10$\,\% linear 
polarization can be reached (for $\Omega_{\rm B} \ge 10\,\Gamma$). 

In unsaturated masers, the largest fractional polarizations observed by us 
($p_{\rm L} \sim 30-40$\,\%) can be reached by means of anisotropic 
pumping (Western \& Watson, \cite{1984ApJ...285..158W}, see however Elitzur,
\cite{1996ApJ...457..415E}, where maser polarization arises in the unsaturated
growth phase). The radiation field that pumps \ch3oh  Class~II masers is 
naturally anisotropic when it comes e.g. from a nearby star, or when the escape 
probability for the pump radiation is anisotropic. As remarked above, the 
latter situation may occur e.g. in a layer compressed by a shock. 
Anisotropic pumping provides a substantial polarization: even 
for $J_{\rm up} > 1$, and $R \sim 10^{-3}\,\Gamma$ (unsaturated regime), and 
without magnetic fields, NW90 find, for axially symmetric pumping, a 
fractional linear polarization of $30-40$\,\%, with a position angle along and 
a maser propagation perpendicular to the symmetry axis of the anisotropic 
pumping. 

For Class~I masers, thought to be collisionally excited (Cragg et al., 
\cite{1992MNRAS.259..203C}), anisotropic pumping is possible if the collisions 
themselves are anisotropic. As suggested by NW90, such conditions may be 
encountered when the collision partners are charged particles undergoing 
ambipolar diffusion in a shock. However, it is uncertain whether the ionization
fractions in maser regions are sufficient to account for anisotropic 
collisional pumping.

In any case, we can also expect polarization due to an anisotropic loss rate 
$\Gamma$, for both maser classes. As mentioned above, the anisotropic 
absorption of IR photons from the maser levels, populating torsionally excited 
states, will imply an unequal population of the magnetic substates of the maser 
levels. 

\subsection{Application to the observed mm \ch3oh  masers} 
It is very likely that most of our \ch3oh  masers are unsaturated, 
given their weakness with respect to the lower frequency transitions of the 
same class. Unfortunately, determining the degree of maser saturation is 
notoriously difficult. In the following, we assume that the observed 
mm \ch3oh  masers are unsaturated. As discussed above, anisotropic pumping 
or anisotropic radiative losses are then a necessary condition to achieve 
the observed linear polarization.

Although a quantitative analysis of \ch3oh  maser polarization is still lacking,
we can discuss our results in view of the models presented in the previous 
section. Figure\,\ref{histograms} shows the distribution of detections of 
fractional polarization among the sources listed in 
Tables\,\ref{maser_survey_class1} and \ref{maser_survey_class2}. The highest 
linear polarizations (35-40\,\%) are reached in both classes. Although our 
sample is still too small to derive reliable statistics from it, we find that 
a large fractional linear polarization is regularly achieved in both classes of 
mm CH$_3$OH masers. Whether anisotropic pumping and anisotropic radiative losses
are needed or not to achieve such polarizations is a matter of current debate.
 
\subsection{Discussion of individual sources}
\subsubsection{\object{OMC-2}}
\begin{figure}
     \resizebox{8.5cm}{!}{\includegraphics[bb=70 30 500 540]{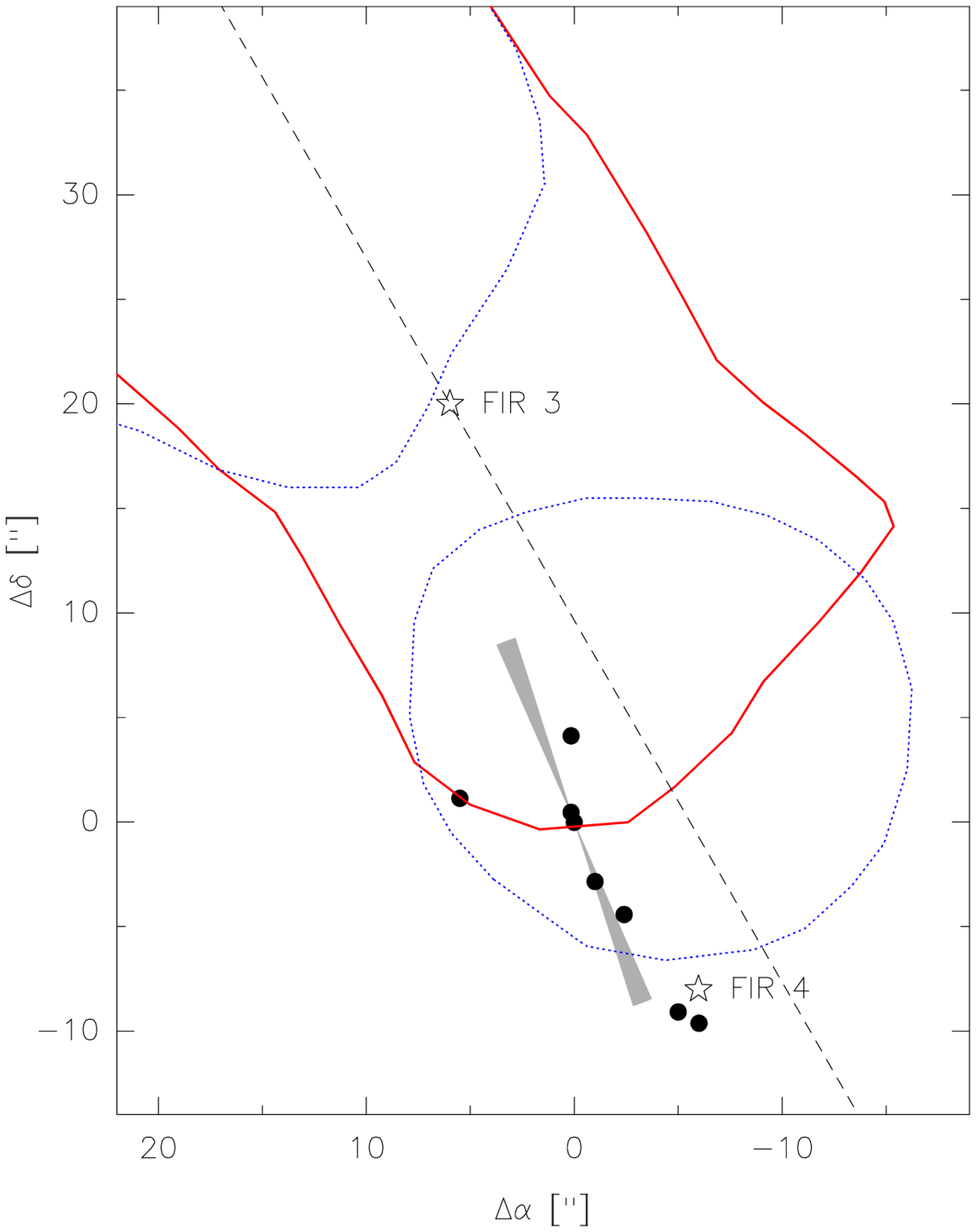}} 
     \caption{Schematic representation of methanol masers in \object{OMC-2}.
     Black dots: 44\,GHz masers of Kogan \& Slysh (\cite{1998ApJ...497..800K}).
     Contours: extent of the overlapping outflow system (in CO(1-0), from 
     Williams et al., \cite{2003ApJ...591.1025W}). The dashed straight line is 
     the outflow axis. Dotted contours: blue shifted lobes. Solid contour: 
     red shifted lobe. The polarization angle of our 133\,GHz measurement and 
     its error are indicated by the shadowed line (whose length is the 
     beamwidth, {\sc fwhm}). The positions of the FIR sources are from Chini et 
     al.  (\cite{1997ApJ...474L.135C}). \label{omc2}}
   \end{figure}

The direction of the linear polarization of the 133\,GHz Class~I maser in 
OMC~2 is close to the outflow axis (Fig.\,\ref{omc2}). The outflow direction is
well defined both by CO observations (Williams et al., 
\cite{2003ApJ...591.1025W}) and by the alignment of the 44\,GHz maser spots 
(Kogan \& Slysh, \cite{1998ApJ...497..800K}). Since a shock at the "working 
surface" of the outflow will be too fast and destroy molecules, an oblique 
shock, propagating away from the outflow axis, explains both the required  
enhanced gas-phase \ch3oh  abundance, and the asymmetry for an anisotropic loss
rate (i.e. escape probability for IR photons). In this scenario, the magnetic 
field tends to line up roughly along the shock front, and only a maser 
propagation parallel to the shock plane will provide a strong maser gain. The 
magnetic field and the maser propagation can then be almost perpendicular, 
because this outflow is not viewed pole-on. Such a model provides an 
explanation for both the observed fractional linear polarization 
($p_{\rm L} = 7\,\%$ despite $R \la \Gamma$) and its orientation parallel to 
${\bf B}_{\rm sky}$ and hence to the outflow axis (cf. NW90, their Fig. 11, 
case (2)). 

\subsubsection{DR21\,-W}
The case of DR21\,-W is less clear. Slysh et al. (\cite{1997ApJ...478L..37S}) 
observed that the 84\,GHz and 133\,GHz lines have similar spectral features. 
All the more it is astonishing that in \object{DR\,21-W}, both masers, similar 
in strength, line shape and fractional linear polarization have polarization 
angles that differ by $40^\circ$. The reason may simply be the variability of 
unsaturated maser emission (the 84\,GHz maser was observed in February 2002, 
the one at 133\,GHz in May 2002). If the polarization characteristics are 
constant in time, and if the spatial separation of the masers at 84\,GHz and 
133\,GHz is similar to that of the two 44\,GHz components of Kogan \& Slysh, 
this could be a hint at an inhomogeneous magnetic field and/or an anisotropic 
pump rate $P$ or loss rate $\Gamma$ with different symmetry axes for both 
masers (for a distance of 3\,kpc, $2\farcs 8$ correspond to 8000 AU, 
cf. Plambeck \& Menten, \cite{1990ApJ...364..555P}, further references therein).
However, the spectral profiles of the polarization angle of either source is 
flat, i.e. the magnetic field and/or $P$ and $\Gamma$ are homogeneous at least 
along each maser's axis. 
   \begin{figure}[hb!]
     \resizebox{8.5cm}{!}{\includegraphics[bb=70 30 500 540]{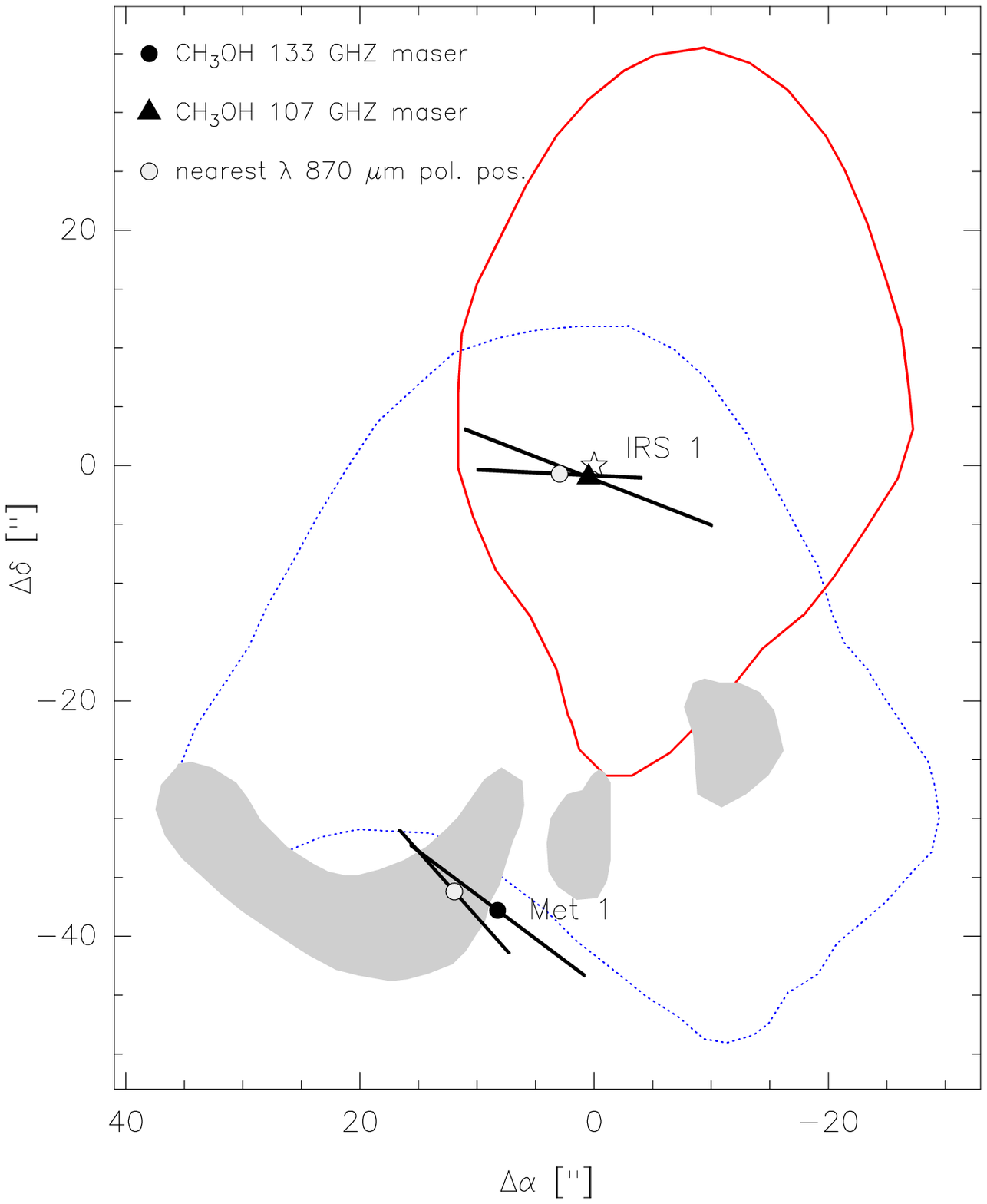}} 
     \caption{Schematic representation of the mm methanol masers in 
     \object{NGC\,7538}, centered at the position of IRS\,1. Contours: extent 
     of the outflow system (in CO(1-0), from Kameya et al., 
     \cite{1989ApJ...339..222K}). Dotted contour: blue shifted lobe. Solid 
     contour: red shifted lobe. The polarization angle of our 107\,GHz \& 
     133\,GHz \ch3oh  maser measurements (filled circle resp. filled triangle) 
     as well as that of the nearest $\lambda\,870\,\mu$m continuum polarization
     measurements (open circles, Momose et al., \cite{2001ApJ...555..855M}) are 
     indicated by the straight lines (whose lengths indicate the beamwidth, 
     {\sc fwhm}). The grey-shaded area indicates the H$_2$ emission suggesting 
     a bow shock (Davis et al., \cite{1998AJ....115.1118D}). \label{ngc7538}}
   \end{figure}

\subsubsection{\object{NGC\,7538}}
In \object{NGC\,7538-Met1}, the linear polarization is roughly perpendicular to
a suggestive bow-shock structure as traced by shock-excited H$_2$ emission
(Davis et al., \cite{1998AJ....115.1118D}). This is schematically shown in 
Fig.\,\ref{ngc7538}. As for \object{OMC-2}, the polarization angle profile is 
rather flat, and a similar scenario may apply. The polarization angles of the 
133\,GHz Class~I maser towards \object{NGC\,7538-Met1} and of the 107\,GHz
Class~II maser towards \object{NGC\,7538-IRS1} are very close to those of the 
$\lambda\,870\,\mu$m continuum emission from these positions. Since the 
magnetic field direction in the dust emitting region and the maser region is 
likely to be the same, this may be taken as evidence for magnetic alignment of 
dust grains. However, in the unsaturated regime, it cannot be ruled out that 
the linear polarization is rather due to the amplification of polarized 
background radiation from dust. Then the fractional linear polarization of the 
maser radiation and its polarization angle are the same as for the continuum 
seed radiation (see e.g.  Goldreich et al., \cite{1973ApJ...179..111G}, eq. 64, 
or Wallin \& Watson, \cite{1997ApJ...481..832W}, equations 17--19), whatever the
explanation of the grain alignment may be. This is only valid if the 
magnetic substates are equally populated, i.e. the maser has to be weak, so 
that its own radiation does not yet affect the absorption coefficients. In order
to fulfill this condition, we have to replace the inequality $R \ll \Gamma$ by 
the more stringent $R/\Gamma \ll \Omega_{\rm B}/\Delta\omega_{\rm D}$, where 
$\Delta\omega_{\rm D}$ is the Doppler linewidth. For a 10\,mG field, and a mean 
maser linewidth of 0.5\,\kms ({\sc fwhm}), we have 
$\Omega_{\rm B}/\Delta\omega_{\rm D} \sim 10^{-5}$, since the \ch3oh  molecule 
is non-paramagnetic. In turn, $R << 10^{-5}\,\Gamma$ is only valid for rather 
weak masers, otherwise the polarization due to the seed radiation will be 
destroyed by the (still unsaturated) maser itself. An observable quantity which
allows to distinguish between both regimes is the line-to-continuum flux ratio.
Since this ratio was only measurable with limited accuracy and spatial 
resolution towards two sources (see section 3), the data presented here do not 
yet allow for a meaningful conclusion. Interestingly, Pestalozzi et al. 
(\cite{2004ApJL...p}) successfully model the 12.2 and 6.7~GHz Class~II methanol 
maser towards \object{NGC\,7538-IRS1} by high amplification of background radio 
continuum emission. 
 
\subsection{Spectral profiles of the polarization angle}
\begin{figure}
   \centering
     \resizebox{\hsize}{!}{\includegraphics{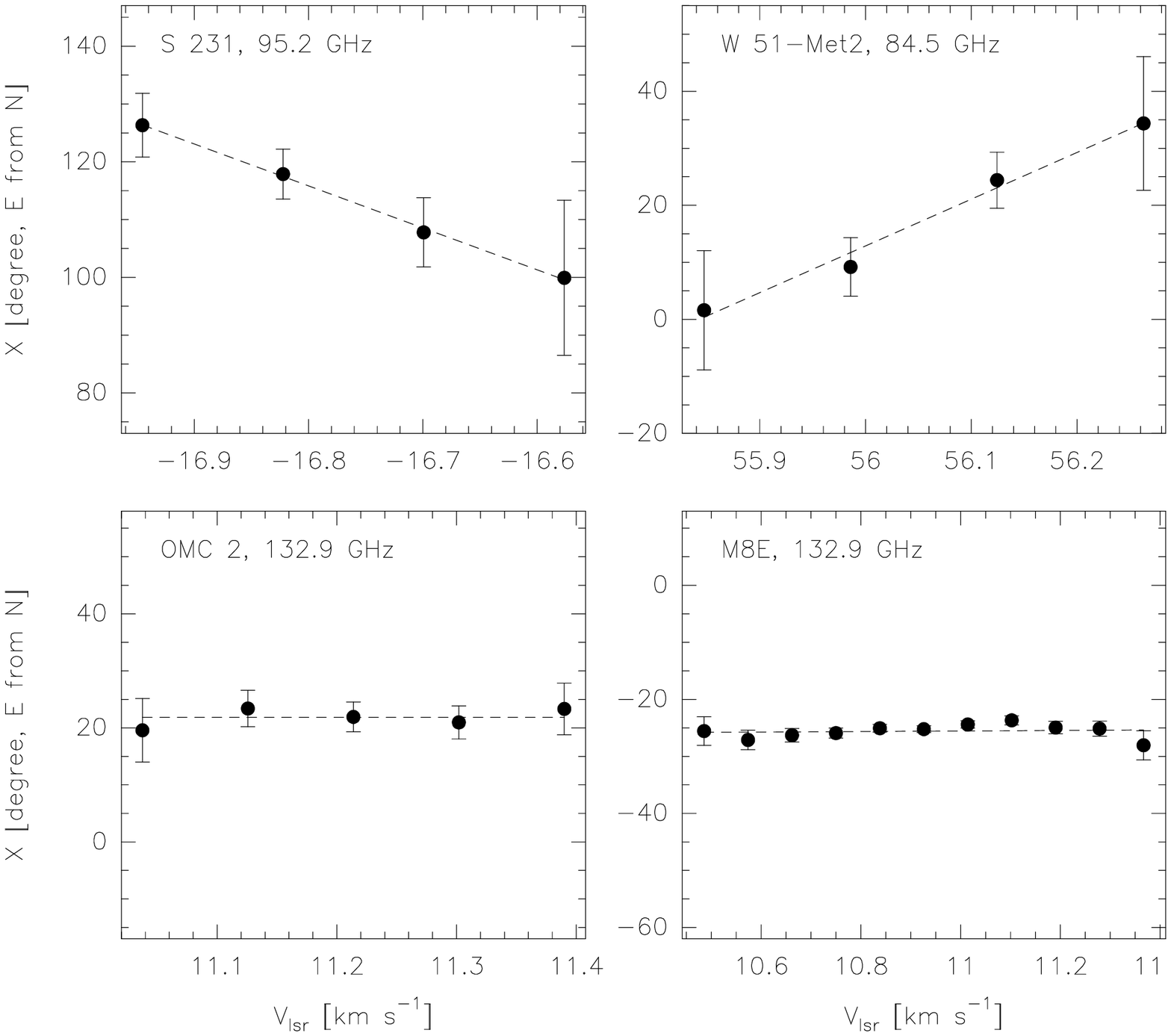}}
     \caption{Polarization angle profiles of Class~I \ch3oh  masers. The 
     ordinates are all at the same scale, to better compare the end-to-end 
     difference of polarization angles. The flat profiles (bottom row) are 
     clearly distinguished from those with a slope (top row).
     \label{chiprofiles}}
   \end{figure}

Most of the polarized Class~I masers have flat polarization angle profiles. Only
two sources (\object{S\,231} and \object{W\,51-Met\,2}) have an almost constant
slope (Fig.\,\ref{chiprofiles}), of $-72\fdg 8/$\kms  and $97\fdg 6/$\kms, 
respectively. As shown by NW90, a slope in the polarization angle profile is 
produced by a gradient in the magnetic field direction, for a constant loss 
rate $\Gamma$ and pump rate $P$. Consequently, the spectral profile of the 
linear polarization has a maximum at the velocity where the angle between the 
magnetic field and the maser propagation direction is largest (clearly visible 
in \object{S\,231} at 95\,GHz, Fig.~\ref{spectra_95ghz}). 

Keplerian disks with an azimuthal, frozen-in magnetic field component
following the velocity vectors (as modelled by Barvainis,
\cite{1984ApJ...279..358B}, for the SiO maser in Orion\,IRc\,2) are an 
attractive model providing an unconstrained explanation for both 
a gradient of the magnetic field direction, and for an anisotropic pump
or loss rate. For the case of the 6.7\,GHz Class~II maser in \object{S\,231}, 
Minier et al. (\cite{2000A&A...362.1093M}) propose a masing disk as explanation 
of the kinematics of the maser spots. Pestalozzi et al. (\cite{2004ApJL...p})
explain the main component of the 12.2 and 6.7\,GHz \ch3oh  maser emission 
towards \object{NGC\,7538-IRS1} by a differentially rotating disk. On the other
hand, De Buizer (\cite{2003MNRAS.341..277D}) casts doubt on the disk
interpretation, finding that the shock-excited H$_2$ emission towards young 
stellar objects associated with linearly distributed methanol masers is mostly 
aligned both with the associated outflows and with the methanol maser spots 
(which should be aligned perpendicularly to the outflow in the disk scenario).
After all, a slope in the polarization angle profile can also simply be due to a
changing direction of the symmetry axis for an anisotropic pump or loss rate,
without any need for a particular magnetic field structure. Only more accurate 
models and higher spatial resolution observations will allow to distinguish 
between those scenarios.
   \begin{figure}
   \centering
     \resizebox{\hsize}{!}{\includegraphics{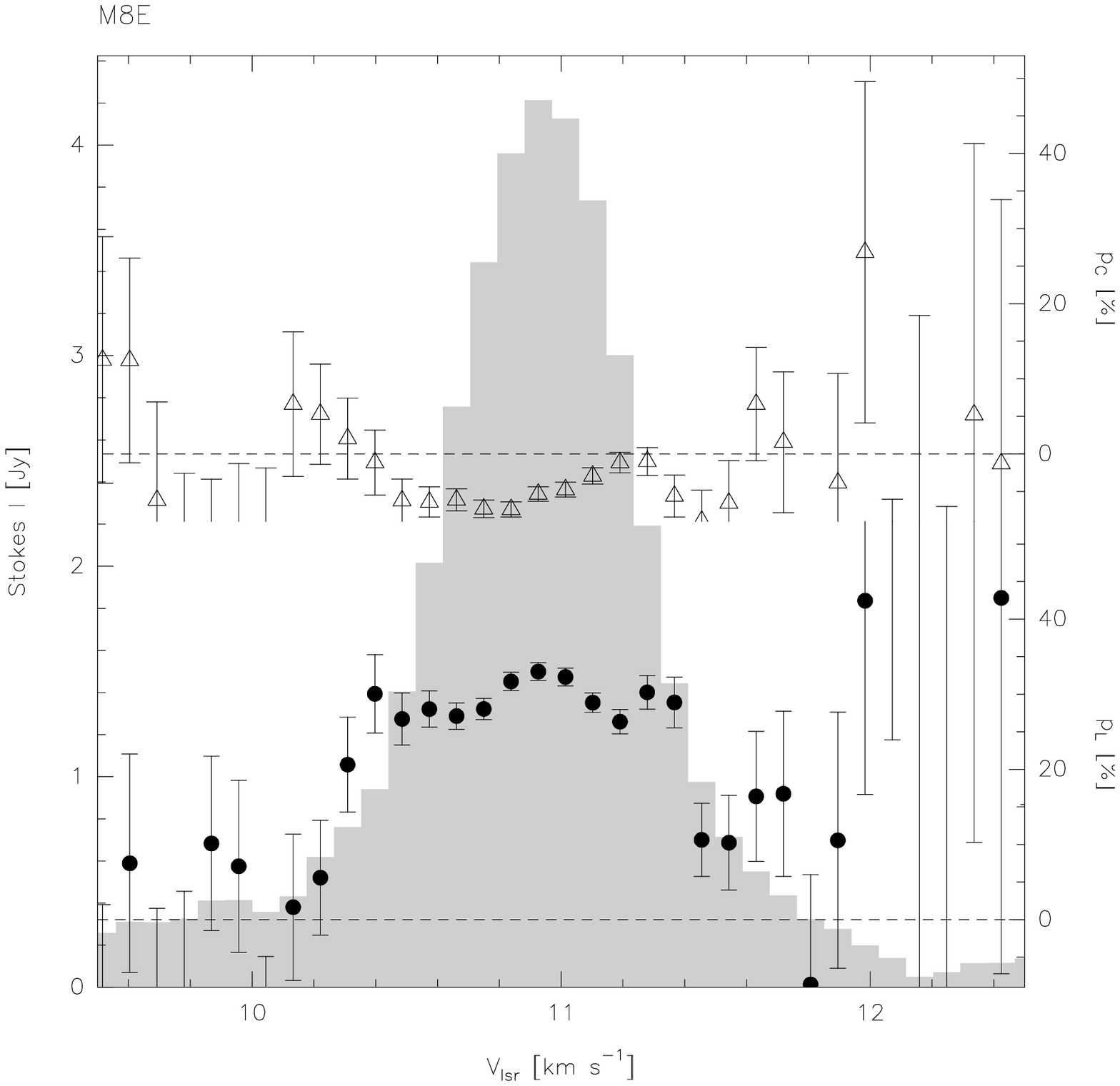}}
     \resizebox{\hsize}{!}{\includegraphics{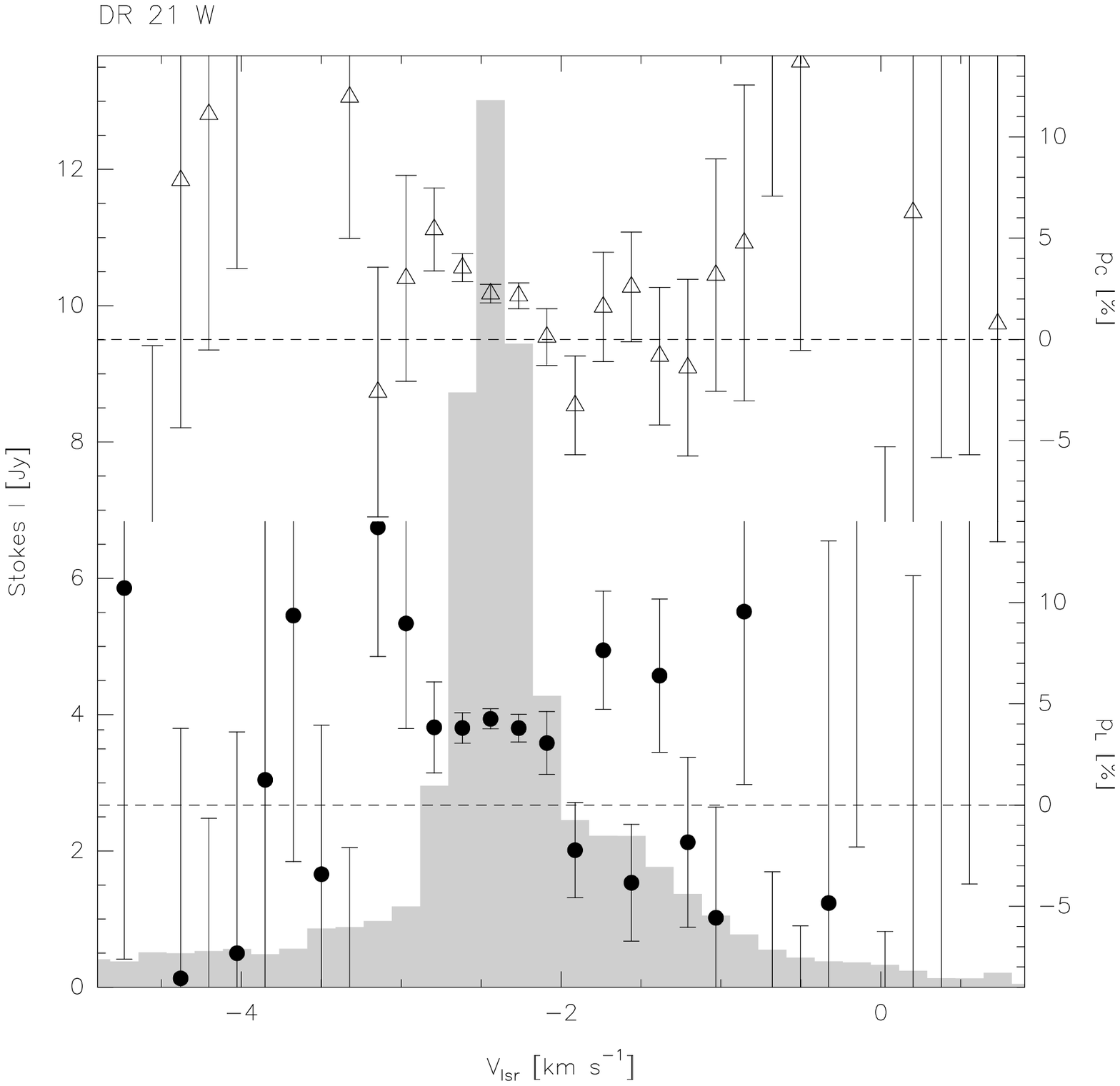}}
     \caption{Detections of circular polarization towards \object{M8E} (upper)
     and \object{DR\,21-W} (lower), both in the 133\,GHz \ch3oh  maser. The 
     circular polarization $p_{\rm C}$ is indicated by open triangles, and the 
     linear polarization $p_{\rm L}$ by filled dots (both on the same scale). 
     The dashed lines indicate the level of zero polarization. For comparison, 
     the Stokes I spectra are shown as grey-shaded histograms. The axes of 
     \object{M8E} and \object{DR\,21-W} are not on the same scale.}
     \label{spectra_pc}
   \end{figure}
\subsection{Circular polarization}
The tentative detection of circular polarization in two Class~I methanol masers
(Fig.~\ref{spectra_pc}) is interesting and the mechanism producing the effect 
is not clear. We note that the Zeeman components of the \ch3oh  maser are 
unresolved\,: As estimated in section 4.5.3, 
$\Omega_{\rm B}/\Delta\omega_{\rm D} \sim 10^{-5}$.
Due to the unsaturated maser emission, the circular
polarization observed in the 133\,GHz masers of \object{M8E} and 
\object{DR\,21-W} can only be due to non-Zeeman effects, a conclusion also 
supported by the absence of the classical S-shaped signature of Stokes V. As 
estimated above, we have $\Omega_{\rm B} \gg R$ and $\Gamma$, and the symmetry 
axis for the molecular quantum substates ("quantization axis") in the masing 
medium is aligned with the magnetic field. Circular polarization in this regime 
was investigated by Wiebe \& Watson (\cite{1998ApJ...503L..71W}), and results 
from a misalignment of the local quantization axis and the direction of linear 
polarization. Such a misalignment can be achieved by changes of 
${\bf B}_{\rm sky}$ along the maser propagation path. Towards Class~II masers, 
we did not detect circular polarization. Our most stringent upper limit is 
$p_{\rm C} < 2.5\,\%$ (for \object{NGC\,7538-IRS1}, 107\,GHz maser). Likewise, 
Koo et al. (\cite{1988ApJ...326..931K}) get a negative result from their survey 
of the 12\,GHz line. 

\subsection{The velocity components of Class~II masers}
\begin{figure}
     \resizebox{\hsize}{!}{\includegraphics{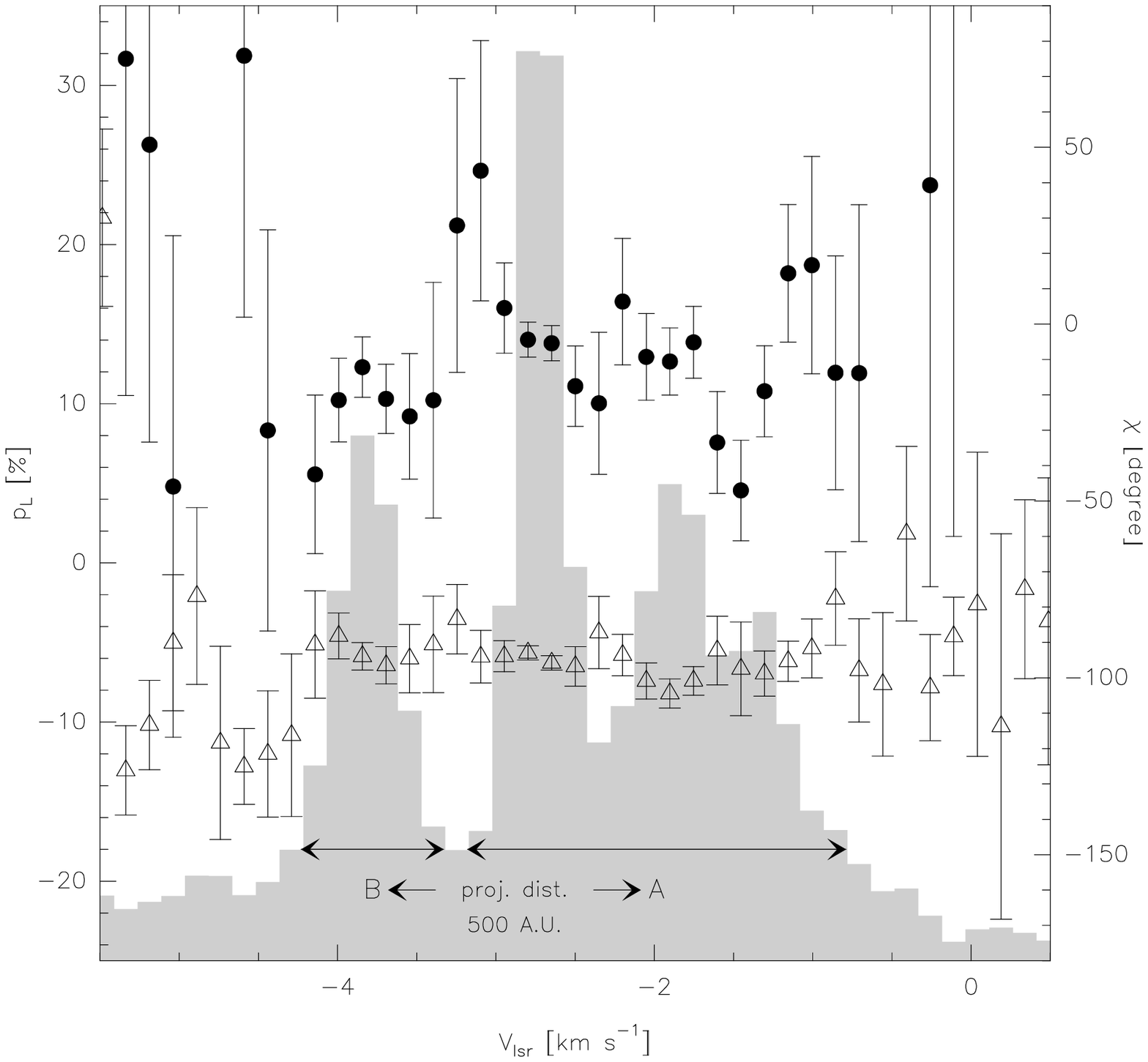}} 
     \caption{Overlay of the linear polarization (filled circles, left ordinate)
     and the polarization angle (unfilled triangles, right ordinate) with the 
     spectrum (grey-shaded) of the 157\,GHz \ch3oh  maser towards 
     \object{Cep\,A}. The projected distance of the components A and B is 
     indicated (derived by Mehringer et al. (\cite{1997ApJ...475L..57M}).
     \label{cepa_zoom}}
   \end{figure}

Another remarkable result is that the different velocity components of a given
Class~II maser, although different in flux density, display 
a similar fractional polarization and polarization angle. This suggests either
masers that are not too far from each other (e.g. 500 A.U. in the case of the 
components A \& B of \object{Cep\,A East}, see Figure~\ref{cepa_zoom}), or that
the velocity of a single maser is redistributed along its propagation direction,
splitting the maser line into different spectral components (e.g. in 
Fig.~\ref{cepa_zoom} the components labelled A in \object{Cep\,A East}, that
are spatially and spectrally unresolved by Mehringer et al., 
\cite{1997ApJ...475L..57M}). In both cases, we can expect a rather homogeneous 
magnetic field structure and/or distribution of the anisotropy of the pumping 
radiation field. As for velocity redistribution, Nedoluha \& 
Watson (\cite{1988ApJ...335L..19N}) show that such a breakup into multiple, 
narrow lines is even possible in presence of a smooth velocity gradient of the 
order of a few \kms  along the maser. Despite these changes in velocity, the 
magnetic field and - perhaps more importantly - the direction of the pumping 
anisotropic radiation field remain constant. A constant anisotropy parameter is
expected if the maser length is small compared to the distance from the source 
of pump radiation.

\section{Conclusions}
1. We find linear polarization in 10 out of 14 Class~I sources, and in 3 out 
of 7 Class~II sources. The highest polarization ($39.5\,\%$) is found in 
the Class~I maser \object{L\,379}, the most significant one in \object{M8E}
($p_{\rm L}/\sigma_{\rm rms} > 25$, both at 133\,GHz). Both Class~I and 
Class~II mm {\ch3oh} masers reach substantial polarizations well above 10\,\%. 
The mm \ch3oh  masers, given their weakness relative to those at lower 
frequencies, are unsaturated. The polarization observed possibly needs, 
especially for the larger angular momenta, anisotropic pumping and an 
anisotropic escape probability leading to an anisotropic loss rate. The maximal 
polarization of Class~I masers is close to that observed for H$_2$O masers. 
None of our three polarized Class~II sources reaches a polarization as high as 
that of SiO masers.
\\[1.5ex]
2. For Class~I masers, the polarization angle profiles across the maser lines 
are mostly flat; two sources (W51-Met\,2 \& \object{S\,231}) show evidence of a
linear slope, suggesting that the direction of the magnetic field and of the 
symmetry axis for anisotropic loss or pumping changes along the propagation 
axis of the maser. The more common flat polarization angle profiles could be 
attributed to masers originating from shocks that are seen edge-on, providing 
the largest possible gain lengths and a highly anisotropic escape probability.
\\[1.5ex]
3. The polarization angles of the Class~I and II masers in \object{NGC\,7538}
show a remarkable agreement with the results from submillimeter continuum 
polarization, the latter being generated by magnetically aligned grains. 
We suggest that the Class~I maser is associated with a bow shock in the 
outflow. The magnetic field in this maser is likely enhanced by compression; 
due to the shock, the loss rate is anisotropic, yielding a 133~GHz maser
with a relatively strong polarization of 24\,\%. The 107\,GHz Class~II maser 
towards IRS1 is weakly polarized. For both masers, the polarization
angles are close to those of the submm continuum towards the respective 
positions of the masers. This provides evidence for magnetic alignment of dust 
grains, the magnetic fields in the dust emitting and masing volumes having the 
same direction. Another explanation is the amplification of the polarized 
continuum emission, which provides the seed photons for this unsaturated maser. 
In several of the polarized Class~I and Class~II mm \ch3oh  masers shown here, 
the line-to-continuum flux ratio is sufficiently small to make such a scenario 
plausible.
\\[1.5ex]
4. The polarization direction of the prominent Class~I maser in \object{OMC-2}
is close to the outflow axis. This observation can be interpreted as a 
maser originating in a lateral C-type shock compressing the gas in a plane 
roughly parallel to the outflow axis, making the loss rate anisotropic.
The linear polarization -- relatively strong for an unsaturated maser --
indicates a maser propagation perpendicularly to the magnetic field. The flat
spectral profile of the polarization angle lends additional support to this 
shock scenario. 
\\[1.5ex]
5. All Class~II masers where we detect polarization have several (at least
two) velocity components with the same polarization characteristics. This is 
indicative of velocity redistribution along the propagation direction of a 
single maser with a velocity gradient, and/or of nearby, but unresolved 
individual masers with a constant magnetic field direction and a constant 
direction for anisotropic pumping or radiative losses.
\\[1.5ex]
6. Significant circular polarization of $-7.1$\,\% and 3.5\,\% was tentatively
detected towards two Class~I masers at 133\,GHz (\object{M8E} respectively 
\object{DR\,21-W}). We attribute this result to a non-Zeeman origin, due to the 
weakness of the Zeeman splitting with respect to the linewidth of the \ch3oh  
emission (the molecule is non-paramagnetic), and because the maser emission is 
unsaturated.
\begin{acknowledgements}
We thank our IRAM colleagues M.\,Torres who built the principal component of 
the IF polarimeter, and  D.\,Morris who participated in the development and 
commissioning of the polarimeter. We are also grateful to our Spanish IRAM 
colleagues S.\,Navarro, who helped us to install the polarimeter, G.\,Paubert, 
who configured the backends for it, and A.\,Sievers and the Pico Veleta staff 
who provided assistance with the observations. We acknowledge the comments of an
anonymous referee.
\end{acknowledgements}

\end{document}